\documentclass[times, twocolumn, authoryear, final]{elsarticle}

\usepackage{framed,multirow}

\RequirePackage{geometry}
\usepackage{amssymb}
\usepackage{amsthm}
\usepackage{amsmath}
\usepackage{subfigure}
\usepackage{hyperref}
\usepackage{algorithm}
\usepackage{algpseudocode}
\usepackage{booktabs}
\usepackage{caption}
\usepackage{mathtools}
\usepackage{supertabular}
\usepackage{amsmath}
\usepackage{enumitem}
\usepackage{url}
\usepackage{xcolor}
\usepackage[vskip=0.2em]{quoting}
\journal{Medical Image Analysis}
\begin{document}

\begin{frontmatter}
%\title{Numerical identification of methodological inconsistencies affecting the evaluations in 100+ papers on retinal vessel segmentation, and a new baseline ranking}
\title{A new baseline for retinal vessel segmentation: Numerical identification and correction of methodological inconsistencies affecting 100+ papers}
\author[1]{Gy\"orgy Kov\'acs\corref{cor1}}
%\fnref{fn2}
%}
\ead{gyuriofkovacs@gmail.com}

\author[2]{Attila Fazekas%
%\fnref{fn1}
}
\ead{attila.fazekas@inf.unideb.hu}

\cortext[cor1]{Corresponding author}
%\fntext[fn1]{This is the first author footnote.}
%\fntext[fn2]{Another author footnote, this is a very long
%footnote and it should be a really long footnote. But this
%footnote is not yet sufficiently long enough to make two
%lines of footnote text.}
\address[1]{Analytical Minds Ltd., \'Arp\'ad street 5, Beregsur\'any 4933, Hungary}
\address[2]{University of Debrecen, Faculty of Informatics, P.O.BOX 400, Debrecen 4002, Hungary}


%\author{Attila \snm{Fazekas}%\corref{cor1}\fnref{1}} 
%\cortext[cor1]{Corresponding author: Tel.: +36-52-512-900}
%\ead{attila.fazekas@inf.unideb.hu}
%\author{Gy\"orgy \snm{Kov\'acs}\fnref{2}}
%\address[1]{University of Debrecen, Faculty of Informatics, P.O.BOX 400, Debrecen 4002, Hungary}
%\address[2]{Analytical Minds Ltd., \'Arp\'ad street 5, Beregsur\'any 4933, Hungary}

%\received{1 May 2013}
%\finalform{10 May 2013}
%\accepted{13 May 2013}
%\availableonline{15 May 2013}
%\communicated{S. Sarkar}

%\received{1 May 2013}
%\finalform{10 May 2013}
%\accepted{13 May 2013}
%\availableonline{15 May 2013}
%\communicated{S. Sarkar}

\begin{abstract}
%As the characteristics of the vasculature in retinal images can indicate the presence of various diseases and facilitate the localization of other anatomical parts, 
In the last 15 years, the segmentation of vessels in retinal images has become an intensively researched problem in medical imaging, with hundreds of algorithms published. One of the \emph{de facto} benchmarking data sets of vessel segmentation techniques is the DRIVE data set. Since DRIVE contains a predefined split of training and test images, the published performance results of the various segmentation techniques should provide a reliable ranking of the algorithms. Including more than 100 papers in the study, we performed a detailed numerical analysis of the coherence of the published performance scores. We found inconsistencies in the reported scores related to the use of the field of view (FoV), which has a significant impact on the performance scores. We attempted to eliminate the biases using numerical techniques to provide a more realistic picture of the state of the art. Based on the results, we have formulated several findings, most notably: despite the well-defined test set of DRIVE, most rankings in published papers are based on non-comparable figures; in contrast to the near-perfect accuracy scores reported in the literature, the highest accuracy score achieved to date is 0.9582 in the FoV region, which is 1\% higher than that of human annotators. The methods we have developed for identifying and eliminating the evaluation biases can be easily applied to other domains where similar problems may arise.
\end{abstract}


%\maketitle

\end{frontmatter}

%\linenumbers

\section{Introduction}

Retinal images provide a noninvasive way to detect and monitor various conditions and diseases of the eye and body, such as diabetes-related retinopathies, macular degeneration, glaucoma and hypertension \citep{diseases}.
Consequently, the computer-aided (semi-)automatic analysis and screening of retinal images has become a highly popular field of study in medical imaging, with thousands of papers and algorithms focusing on various anatomical parts and degenerations observable in retinal images \citep{diseases, algorithms}.
The popularity of the field can be illustrated by the numerous challenges on the segmentation and grading of retinal images in recent years, organized by data science competition platforms (such as the Diabetic Retinopathy Challenge on Kaggle\footnote{\url{https://www.kaggle.com/c/diabetic-retinopathy-detection}}), and as special tracks of corresponding conferences (like "IDRiD: Diabetic Retinopathy – Segmentation and Grading Challenge" \citep{idrid}).

\begin{figure}[htb]
     \subfigure[\label{fundus}]{\includegraphics[width=0.23\textwidth]{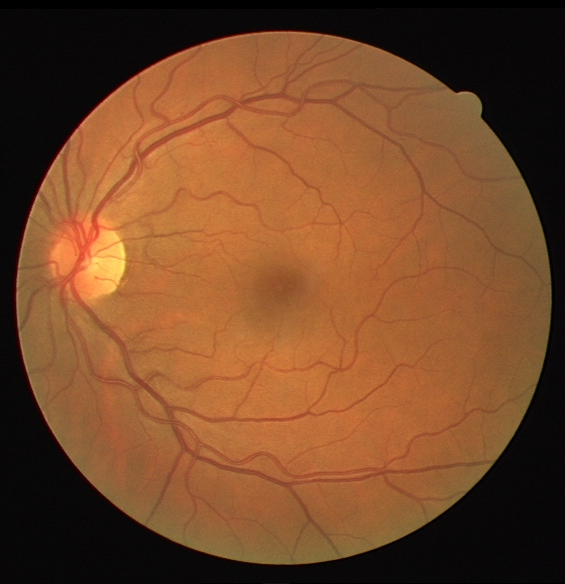}}
     \subfigure[\label{manual1}]{\includegraphics[width=0.23\textwidth]{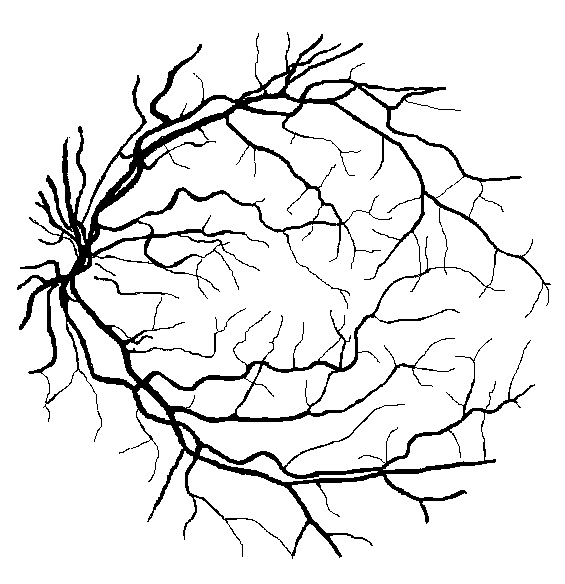}}\\
     \subfigure[\label{manual2}]{\includegraphics[width=0.23\textwidth]{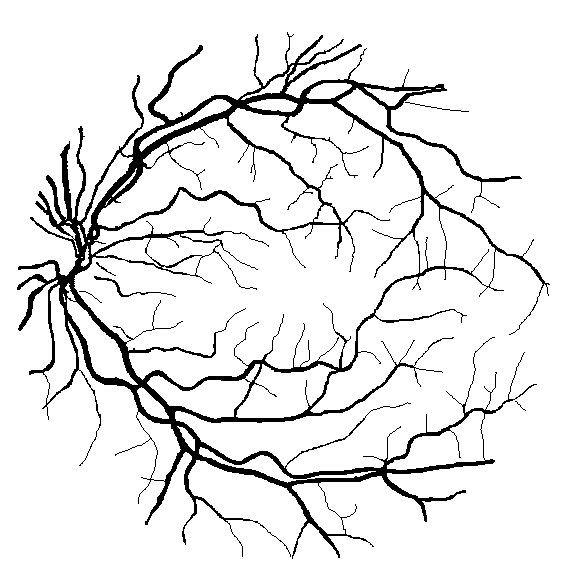}}
     \subfigure[\label{mask}]{\includegraphics[width=0.23\textwidth]{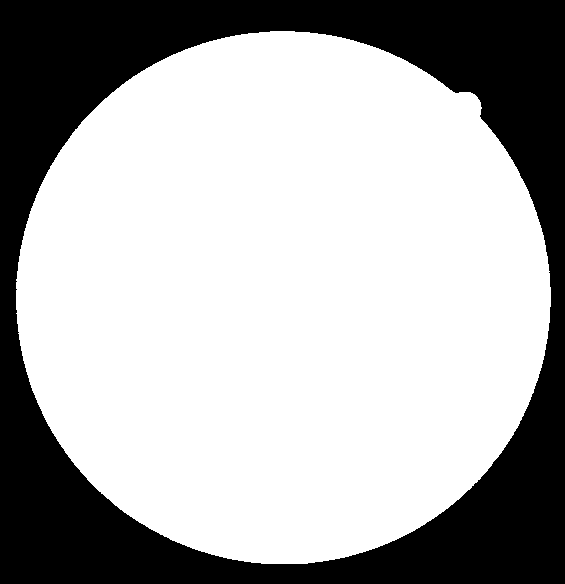}}
     \caption{The "01" entry of the test set in the DRIVE database: retinal image (a); manual segmentation \#1 (with inverted intensities for visibility) (b); manual segmentation \#2 (inverted) (c); field-of-view (FoV) mask, covering the meaningful image content (d).}
     \label{fig1}
\end{figure}

A particularly popular problem in retinal image analysis is the segmentation of retinal vessels (illustrated in Figure \ref{fig1}). The importance of the problem is threefold.
On the one hand, the characteristics of the vasculature can aid the early detection of diseases like hypertension \citep{hypertension}.
On the other hand, the removal of vessels usually precedes the detection of other anatomical parts. For example, the high curvature and junction points of thin vessels have similar visual features to microaneurysms \citep{vesselremoval}, consequently, the segmentation of the vasculature can indirectly aid the detection of mycroaneurysms.
Finally, the segmented vasculature may serve as a guide for locating other anatomical parts, such as the macula \citep{localization}.

In competitive research fields, where the primary goal is to find the most accurate solution to a problem, the proper comparison of algorithms is crucial. Comparability requires the coordination of evaluation data sets and methodologies that fit the purpose of the problem. As pointed out recently \citep{additional}, even the grand challenges of recent years tend not to produce robust and reliable rankings. Therefore, the study of evaluations in active research areas such as retinal vessel segmentation is desirable to understand the state of the art.

In a self-organised fashion, three data sets have emerged as the \emph{de facto} standard for the evaluation of vessel segmentation: DRIVE \citep{drive}, STARE \citep{stare} and HRF \citep{budai2013}. Of these, HRF is relatively new, and many classical algorithms have not yet been evaluated on it. STARE is not split into training and test sets, and the ambiguities of cross-validation lead to uncertain and hardly comparable results. In contrast, DRIVE eliminates many ambiguities of the evaluation by providing a predefined split of training and test images. Because of their long history, DRIVE and STARE continue to be the primary data sets for evaluation in recent work \citep{jin2019, pan2019}, and due to predefined test set in DRIVE, the results obtained on DRIVE are generally accepted as comparable, which is reflected in the numerous rankings in the papers showing the merits of the newly proposed algorithms. The most commonly reported performance measures are the accuracy, sensitivity and specificity of the segmentation.

In a previous paper of ours \citep{kovacs2016}, we briefly pointed out that despite the predefined training and test sets in DRIVE, there is a potential methodological flaw in the evaluations, invalidating many rankings and comparisons in the literature. Namely, textual evidence suggests that some authors compute the performance scores in the field-of-view (FoV) region (Figure \ref{mask}), while others compute the scores for all pixels of the images (the entire rectangular area including the black surroundings of the FoV in Figure \ref{mask}). This minor difference affects the accuracy and specificity scores enormously. Due to the continued interest in this topic, we found it worthwhile to investigate the evaluation methodologies and their impact on comparability in detail.
In this paper, we carry out a detailed numerical analysis of the phenomenon including more than 100 papers from the literature. In order to work with a sufficiently large population and eliminate the ambiguity of cross-validation, we investigate the accuracy, sensitivity and specificity scores reported for the test set of DRIVE. Our main contributions to the field can be summarized as follows:
\begin{enumerate}[noitemsep, topsep=0pt]
\item We introduce numerical approaches to uncover some details of the evaluation methodologies used in 100 relevant papers. The results provide inevitable evidence for two main types of methodologies with non-comparable performance scores: evaluation under the FoV mask and evaluation using all pixels.
\item We attempt to eliminate the biases using numerical techniques to produce a more realistic ranking of algorithms.
\item Our results confirm that most rankings in the literature are based on non-comparable figures, the state of the art performance cannot be deduced without the numerical inference and categorization of evaluation methodologies and adjustments of published figures. 
\item The techniques we introduce can be transferred to other areas where the identified methodological flaw is potentially present.
\item According to our best knowledge, this is the first study investigating the coherence of a subfield of medical image analysis by numerical techniques.
\item For complete reproducibility, the raw data, the implementation of the analysis and the results are shared in the GitHub repository \url{https://github.com/gykovacs/retina_vessel_segmentation}.
\end{enumerate}

The paper is organized as follows. In Section \ref{sec-materials}, we describe the details of the data set DRIVE and formulate the problem we address by providing textual evidence for its existence. In Section \ref{sec-analysis}, we introduce the numerical methods and present the raw results. Based on the findings of the numerical analysis, we discuss the state of the art in Section \ref{overall}. Finally, we summarize the paper and draw conclusions in Section \ref{sec-discussion}.

\section{Problem formulation}
\label{sec-materials}

In this section, we describe the main characteristics of the data set DRIVE, the performance scores, and we also discuss and illustrate the root cause of the problem we address.

\subsection{The DRIVE data set}

The Digital Retinal Images for Vessel Extraction (DRIVE) data set \citep{drive} contains 20 training and 20 test entries. Images were captured with a Canon CR5 non-mydriatic 3CCD camera with a 45 degree field of view and cropped to 584$\times$565 pixels. Each training entry consists of a retinal image, a binary annotation of vessels (by annotator \#1), and a binary mask of the Field of View (FoV) indicating the useful image content. The FoV masks cover 69\% of the rectangular images. Each test entry includes an additional manual segmentation by annotator \#2, enabling the evaluation of human annotators as a baseline. A complete entry from the test set is visualized in Figure \ref{fig1}. 

\subsection{Performance metrics}

There are numerous metrics in the literature to characterize the performance of image segmentation algorithms. 
In retinal vessel segmentation, most authors report some metrics of binary classification performance at the pixel level. Particularly, a segmentation is compared to the ground truth image to determine the number of
\begin{itemize}[noitemsep, topsep=0pt]
    \item true positives ($tp$) - correctly identified vessel pixels;
    \item true negatives ($tn$) - correctly identified non-vessel pixels;
    \item false positives ($fp$) - non-vessel pixels identified as vessel;
    \item false negatives ($fn$) - vessel pixels identified as non-vessel;
\end{itemize}
and introducing $p=tp + fn$ and $n=tn +fp$ for the total number of vessel and non-vessel pixels, respectively, the scores
\begin{itemize}[noitemsep, topsep=0pt]
\item accuracy ($acc=(tp + tn)/(p + n)$) - the proportion of all correctly identified pixels to all pixels;
\item sensitivity ($sens= tp/p$) - the proportion of correctly identified vessel pixels to all vessel pixels;
\item specificity ($spec= tn/n$) - the proportion of correctly identified non-vessel pixels to all non-vessel pixels
\end{itemize}
are reported at the image level or averaged on all test images. Some authors report the complements of the $sens$ and $spec$ scores known as false negative (1 - $sens$) and false positive (1 - $spec$) rates, respectively.
We highlight that vessel segmentation as a binary classification problem is highly imbalanced, i.e., the number of vessel pixels in the FoV region is 15\% of the number of non-vessel pixels, leading to high absolute accuracy and specificity scores: a 0.1\% improvement in accuracy implies the correct segmentation of about 250 additional pixels, which is 1\% of the size of the vasculature. 

Since the goal is usually to maximize both sensitivity and specificity (and thus maximize accuracy), some authors report other composite measures of sensitivity and specificity, such as the F$_1$-score (the harmonic mean of $sens$ and $spec$) or the AUC score \citep{rocauc}, however, these scores are reported too infrequently to rely a comprehensive analysis on them. Therefore, in the remainder of the paper, we focus on the $acc$, $sens$ and $spec$ scores. 

\subsection{The source of ambiguity}
\label{ambig}
Although DRIVE specifies dedicated training and test sets, there is some ambiguity regarding the comparability of published performance scores. 
For example, one of the earliest papers \citep{staal2004} explicitly states:
\begin{quoting}
"Since the dark background outside the FOV is easily extracted, in this paper all experiments are done on the FOV only."
\end{quoting}
On the other hand, \citep{villalobos-castaldi2010} says:
\begin{quoting}
"Based on a total of 329,960 pixels contained in a fundus image of size 565 $\times$ 584
pixels, it is possible to obtain the TP, TN, FP, and FN quantities of the segmented image."
\end{quoting}
Finally, the authors of \citep{yu2012} explicitly state that
\begin{quoting}
"It is mentionable that the accuracy and FPR [false positive rate] measures in Jiang \citep{jiang2003}, Martinez-Perez \citep{martinez-perez2007}, and Miri's \citep{miri2011} methods were calculated in the complete image instead of the FOV."
\end{quoting}

Even this handful of quotations suggests that some authors evaluate performance only in the area covered by the FoV mask, while others use all pixels of the images. 
The pixels outside the FoV are usually non-vessel pixels accounting for about 31\% of all pixels in the images. If they are treated as true negatives, the accuracy and specificity increases tremendously. To illustrate the magnitude of this effect, we evaluate the segmentation of annotator \#2 against annotation \#1 (as ground truth) under the FoV mask and including all pixels of the images. With the FoV mask, $acc = 0.9473$, $sens= 0.7760$ and $spec=0.9725$, but using all pixels, $acc = 0.9636$, $sens = 0.7756$ and $spec=0.9818$. 
Omitting the FoV mask increased the $acc$ and $spec$ values by more than $1.5$\% and $1\%$, respectively. The slight change in sensitivity is due to an average of 15 vessel pixels in the annotations \#1 being outside the FoV. Since many authors report 0.1\% improvements in accuracy, the variation in the region used for evaluation can cause an enormous bias when algorithms are compared and ranked.

In the rest of the paper, we develop numerical techniques to test the consistency of the published figures with the hypotheses that the authors either evaluated under the FoV mask or used all pixels of the images.

\subsection{The need for the numerical analysis}

We attempt to investigate the evaluation methodology of dozens of complex image processing pipelines. Collecting original implementations covering 15 years of technological development seems to be unmanagable. On the other hand, reimplementations of the algorithms are unlikely to reproduce the exact figures shared by the authors as all details of complex image processing pipelines are rarely shared due to length limitations. To avoid further bias and to treat all papers similarly and equally, we decided to base the study only on the shared performance scores. In order to draw conclusions without statistical uncertainties, we decided to rely on numerical analysis to reveal some details of the evaluation methodologies used by the authors.

\section{Methods and raw results}
\label{sec-analysis}

In this section, the steps of the numerical analysis are described and the corresponding results are reported. The raw results are turned into insights and discussed in detail in Section \ref{overall}. We introduce two special notations: The mean of a set of figures is denoted by an overline (such as $\overline{acc}$ standing for the mean of a set of accuracy scores), and asterisk indicates exact values with no uncertainty (such as $tp^*$ standing for the exact number of true positives) that can either be extracted from the publicly available data set or denote unknown but exact values extracted by the authors to calculate the performance scores they reported.

\subsection{The papers and algorithms involved}

To select a sufficiently large, influential and comprehensive population of algorithms and reported performance scores without subjective bias, we applied the following strategy. We searched for the term \emph{retinal vessel segmentation} in three of the largest scientific publication databases (\emph{Elsevier ScienceDirect}, \emph{IEEE Xplore} and \emph{ACM Digital Library}) and recursively added the most cited ones with all algorithms referenced in them until we reached 90 papers. Finally, 10 recent papers, mainly based on deep learning have been added to represent the latest trends. To maintain quality and achieve the goals of our study, papers that met any of the following exclusion criteria were omitted in the selection process:
\begin{itemize}[noitemsep,topsep=0pt]
    \item not sharing $acc$, $sens$ and $spec$ scores for DRIVE;
    \item early papers of authors with multiple papers in the field;
    \item being published only on preprint servers;
    \item not reporting performance scores to at least 3 decimal places (rounding of the scores to 2 decimal places introduces larger uncertainty than the size of the bias we investigate). 
\end{itemize}

\begin{table*}
\begin{footnotesize}
\begin{center}
\begin{tabular}{l@{\hspace{2pt}}l@{\hspace{2pt}}l@{\hspace{2pt}}l@{\hspace{2pt}}r@{\hspace{2pt}}l@{\hspace{2pt}}l@{\hspace{2pt}}l@{\hspace{2pt}}l@{\hspace{2pt}}@{\hspace{6pt}}l@{\hspace{2pt}}l@{\hspace{2pt}}l@{\hspace{2pt}}l@{\hspace{2pt}}r@{\hspace{2pt}}l@{\hspace{2pt}}l@{\hspace{2pt}}l@{\hspace{2pt}}l@{\hspace{2pt}}}
\toprule
\rotatebox{90}{Rank} &                              Key & $\overline{acc}$ & \rotatebox{90}{Num. image level} &  \rotatebox{90}{Num. aggregated} & \rotatebox{90}{Annotator \#2 acc.} & \rotatebox{90}{Region of eval.} & \rotatebox{90}{Decimal places} & \rotatebox{90}{Citations} & \rotatebox{90}{Rank} &                           Key & $\overline{acc}$ & \rotatebox{90}{Num. image level} &  \rotatebox{90}{Num. aggregated} & \rotatebox{90}{Annotator \#2 acc.} & \rotatebox{90}{Region of eval.} & \rotatebox{90}{Decimal places} & \rotatebox{90}{Citations} \\
\midrule
                   1 &             \cite{jebaseeli2019} &            .9898 &                                  &                                1 &                                    &                               F &                              4 &                         6 &                   51 &               \cite{meng2015} &            .9529 &                               20 &                                1 &                              .9470 &                               F &                              4 &                        13 \\
                   2 &                \cite{hassan2018} &            .9793 &                               20 &                                1 &                                    &                                 &                              4 &                        24 &                   52 &                 \cite{li2016} &            .9527 &                                2 &                                3 &                              .9472 &                                 &                              4 &                       306 \\
                   3 &                  \cite{wang2015} &            .9767 &                               20 &                                9 &                              .9464 &                               F &                              4 &                       274 &                   53 &              \cite{imani2015} &            .9523 &                               20 &                                5 &                              .9473 &                                 &                              4 &                        84 \\
                   4 &   \cite{villalobos-castaldi2010} &            .9759 &                                  &                                1 &                              .9473 &                               a &                              4 &                        70 &                   54 &              \cite{singh2016} &            .9522 &                               20 &                                1 &                                    &                               a &                              4 &                        75 \\
                   5 &                \cite{memari2017} &            .9722 &                                  &                                3 &                              .9464 &                                 &                              4 &                        23 &                   55 &                 \cite{mo2017} &            .9521 &                                2 &                                3 &                                    &                                 &                              4 &                        68 \\
                   6 &                  \cite{park2020} &            .9706 &                                  &                                1 &                                    &                                 &                              4 &                         2 &                   56 &            \cite{rahmani2020} &            .9521 &                                  &                                1 &                                    &                                 &                              4 &                         0 \\
                   7 &                 \cite{jiang2019} &            .9706 &                                  &                                9 &                              .9637 &                                 &                              4 &                         9 &                   57 &               \cite{dash2020} &             .952 &                               20 &                                1 &                                    &                                 &                              3 &                         0 \\
                   8 &                  \cite{atli2020} &            .9689 &                                  &                                2 &                              .9637 &                                 &                              4 &                         3 &                   58 &          \cite{chalakkal2017} &            .9518 &                                  &                                1 &                                    &                                 &                              4 &                        12 \\
                   9 &            \cite{moghimirad2012} &            .9659 &                                4 &                                1 &                              .9473 &                               F &                              4 &                        44 &                   59 &              \cite{singh2017} &            .9513 &                               20 &                                1 &                                    &                               a &                              4 &                         3 \\
                  10 &                   \cite{pan2019} &            .9650 &                                  &                                1 &                                    &                                 &                              4 &                        12 &                   60 &             \cite{mapayi2015} &            .9511 &                                  &                                6 &                              .9473 &                                 &                              4 &                        50 \\
                  11 &    \cite{escorcia-gutierrez2020} &             .964 &                               20 &                                1 &                                    &                                 &                              3 &                         0 &                   61 &              \cite{zhang2018} &            .9504 &                                  &                                1 &                              .9472 &                                 &                              4 &                        53 \\
                  12 &              \cite{sreejini2015} &            .9633 &                                  &                                1 &                              .9470 &                                 &                              4 &                        47 &                   62 &            \cite{bharkad2017} &            .9503 &                               20 &                                1 &                                    &                                 &                              4 &                         3 \\
                  13 &                 \cite{saleh2011} &            .9630 &                                  &                                2 &                              .9473 &                                 &                              4 &                        68 &                   63 &            \cite{barkana2017} &            .9502 &                                4 &                                4 &                              .9470 &                               F &                              4 &                        72 \\
                  14 &                 \cite{kumar2016} &            .9626 &                                  &                                1 &                                    &                                 &                              4 &                        28 &                   64 &               \cite{khan2016} &            .9501 &                               20 &                                1 &                              .9473 &                                 &                              4 &                        18 \\
                  15 &              \cite{wankhede2015} &            .9626 &                                  &                                1 &                              .9473 &                               a &                              4 &                        19 &                   65 &               \cite{song2017} &            .9499 &                                  &                                1 &                              .9473 &                                 &                              4 &                        12 \\
                  16 &                \cite{pandey2016} &            .9623 &                                  &                                1 &                                    &                                 &                              4 &                        24 &                   66 &             \cite{kovacs2016} &            .9494 &                                  &                                5 &                              .9473 &                               F &                              4 &                        36 \\
                  17 &            \cite{narkthewan2019} &            .9617 &                               20 &                                1 &                                    &                                 &                              4 &                         2 &                   67 &       \cite{roychowdhury2015} &            .9494 &                                  &                                1 &                                    &                                 &                              4 &                       153 \\
                  18 &                \cite{waheed2015} &            .9616 &                               20 &                                1 &                                    &                                 &                              4 &                        23 &                   68 &           \cite{brancati2018} &             .949 &                                  &                                5 &                               .947 &                               F &                              3 &                         0 \\
                  19 &                 \cite{xiang2014} &            .9613 &                                  &                                1 &                              .9612 &                                 &                              4 &                         7 &                   69 &             \cite{nazari2013} &            .9481 &                                  &                                2 &                              .9470 &                                 &                              4 &                        12 \\
                  20 &                  \cite{alom2019} &            .9613 &                                  &                                5 &                                    &                                 &                              4 &                        46 &                   70 &               \cite{kaur2017} &            .9480 &                                  &                                1 &                                    &                                 &                              4 &                        12 \\
                  21 &                  \cite{tang2017} &            .9611 &                               20 &                                1 &                              .9470 &                                 &                              4 &                         6 &                   71 &          \cite{palanivel2020} &            .9480 &                                  &                                4 &                              .9470 &                                 &                              4 &                         1 \\
                  22 &                \cite{samuel2019} &            .9609 &                                  &                                4 &                              .9470 &                                 &                              4 &                        14 &                   72 &              \cite{fraz2012b} &            .9480 &                                2 &                                3 &                              .9464 &                               F &                              4 &                       493 \\
                  23 &                 \cite{tamim2020} &            .9607 &                               20 &                                1 &                                    &                                 &                              4 &                         1 &                   73 &               \cite{shah2017} &            .9479 &                                  &                                1 &                              .9473 &                                 &                              4 &                        18 \\
                  24 &                   \cite{zhu2016} &            .9607 &                               20 &                                9 &                                    &                                 &                              4 &                        92 &                   74 &              \cite{zhang2016} &            .9476 &                                  &                                2 &                              .9472 &                               F &                              4 &                       171 \\
                  25 &             \cite{thangaraj2017} &            .9606 &                               20 &                                1 &                                    &                                 &                              4 &                        16 &                   75 &             \cite{shukla2020} &            .9476 &                                  &                                1 &                              .9473 &                                 &                              4 &                         3 \\
                  26 &               \cite{lupascu2016} &            .9606 &                               40 &                                2 &                              .9473 &                                 &                              4 &                         1 &                   76 &              \cite{cheng2014} &            .9474 &                                  &                                6 &                              .9470 &                                 &                              4 &                        80 \\
                  27 &                   \cite{fan2017} &             .960 &                                  &                                2 &                                    &                                 &                              3 &                         7 &                   77 &               \cite{zhou2017} &            .9469 &                                  &                                5 &                              .9473 &                               F &                              4 &                        33 \\
                  28 &               \cite{lupascu2010} &            .9597 &                                7 &                                1 &                              .9473 &                                 &                              4 &                       329 &                   78 &          \cite{melinscak2015} &            .9466 &                                  &                                1 &                                    &                                 &                              4 &                       118 \\
                  29 &                 \cite{ricci2007} &            .9595 &                                2 &                                1 &                              .9473 &                               F &                              4 &                       769 &                   79 &           \cite{mendonca2006} &            .9463 &                                  &                                2 &                              .9473 &                               F &                              4 &                       983 \\
                  30 &                    \cite{wu2020} &            .9582 &                                  &                                3 &                                    &                                 &                              4 &                         9 &                   80 &             \cite{rezaee2017} &            .9463 &                                  &                                1 &                              .9473 &                                 &                              4 &                        34 \\
                  31 &                 \cite{fathi2013} &            .9581 &                               20 &                                1 &                                    &                                 &                              4 &                        95 &                   81 &             \cite{rahebi2014} &            .9461 &                               20 &                                1 &                                    &                                 &                              4 &                        38 \\
                  32 &                  \cite{dash2018} &             .957 &                               20 &                                1 &                                    &                                 &                              3 &                        14 &                   82 &               \cite{miri2011} &            .9458 &                                  &                                1 &                                    &                                 &                              4 &                       291 \\
                  33 &                 \cite{budai2013} &             .957 &                                  &                                1 &                               .947 &                                 &                              3 &                       207 &                   83 &       \cite{strisciuglio2016} &            .9454 &                                  &                                5 &                                    &                                 &                              4 &                        64 \\
                  34 &                   \cite{noh2019} &            .9569 &                                  &                                1 &                              .9472 &                                 &                              4 &                        18 &                   84 &              \cite{marin2011} &            .9452 &                               20 &                                1 &                                    &                               F &                              4 &                       906 \\
                  35 &                \cite{soomro2019} &             .956 &                                  &                                1 &                                    &                                 &                              3 &                        17 &                   85 &              \cite{adapa2020} &            .9450 &                               20 &                                1 &                                    &                                 &                              4 &                         4 \\
                  36 &                \cite{frucci2017} &             .956 &                                  &                                1 &                                    &                                 &                              3 &                         3 &                   86 &             \cite{javidi2017} &            .9450 &                                  &                                1 &                              .9473 &                               F &                              4 &                        46 \\
                  37 &                \cite{frucci2016} &             .955 &                               20 &                                1 &                                    &                                 &                              3 &                        27 &                   87 &          \cite{azzopardi2014} &            .9442 &                                  &                                3 &                                    &                                 &                              4 &                       445 \\
                  38 &                 \cite{saroj2020} &            .9544 &                               20 &                                1 &                                    &                                 &                              4 &                         1 &                   88 &       \cite{strisciuglio2015} &            .9442 &                                  &                                1 &                                    &                                 &                              4 &                        25 \\
                  39 &                   \cite{yan2018} &            .9542 &                                  &                                2 &                              .9472 &                                 &                              4 &                        92 &                   89 &              \cite{staal2004} &            .9441 &                                  &                                1 &                              .9473 &                               F &                              4 &                      2830 \\
                  40 &                  \cite{zhao2015} &             .954 &                                  &                                5 &                               .947 &                                 &                              3 &                       246 &                   90 &                \cite{you2011} &            .9434 &                                  &                                1 &                              .9473 &                               F &                              4 &                       325 \\
                  41 &                    \cite{na2018} &             .954 &                                  &                                1 &                               .947 &                                 &                              3 &                         6 &                   91 &             \cite{soomro2017} &            .9432 &                                  &                                1 &                                    &                                 &                              4 &                        21 \\
                  42 &                 \cite{panda2016} &            .9539 &                                  &                                4 &                              .9470 &                               a &                              4 &                        32 &                   92 &              \cite{kumar2020} &            .9432 &                               20 &                                1 &                              .9473 &                               F &                              4 &                         0 \\
                  43 &          \cite{geetharamani2016} &            .9536 &                               20 &                                1 &                              .9470 &                                 &                              4 &                        77 &                   93 &               \cite{fraz2012} &            .9430 &                                2 &                               11 &                              .9473 &                               F &                              4 &                       294 \\
                  44 &             \cite{liskowski2016} &            .9535 &                                  &                               12 &                              .9473 &                               F &                              4 &                       492 &                   94 &                \cite{dai2015} &            .9418 &                                  &                                1 &                                    &                                 &                              4 &                        41 \\
                  45 &                \cite{soomro2018} &            .9534 &                                  &                                4 &                                    &                                 &                              4 &                        45 &                   95 &           \cite{anzalone2008} &            .9418 &                                  &                                1 &                              .9473 &                               F &                              4 &                        57 \\
                  46 &              \cite{dasgupta2017} &            .9533 &                                  &                                1 &                                    &                                 &                              4 &                       133 &                   96 &         \cite{dizdaroglu2012} &            .9412 &                                  &                                2 &                                    &                                 &                              4 &                        25 \\
                  47 &                   \cite{ngo2017} &            .9533 &                                  &                                1 &                                    &                                 &                              4 &                        19 &                   97 &   \cite{salazar-gonzalez2014} &            .9412 &                                  &                                1 &                              .9473 &                               F &                              4 &                       222 \\
                  48 &                    \cite{hu2018} &            .9533 &                                  &                                4 &                              .9470 &                               F &                              4 &                        87 &                   98 &              \cite{emary2014} &             .939 &                               20 &                                1 &                                    &                                 &                              3 &                        26 \\
                  49 &                  \cite{yang2020} &            .9532 &                                  &                                4 &                                    &                                 &                              4 &                         0 &                   99 &              \cite{zhang2010} &            .9382 &                                  &                                1 &                              .9473 &                               F &                              4 &                       530 \\
                  50 &                 \cite{zhao2015b} &             .953 &                                  &                               10 &                               .947 &                                 &                              3 &                        91 &                  100 &         \cite{odstrcilik2013} &            .9340 &                               20 &                                1 &                              .9473 &                               a &                              4 &                       266 \\
\bottomrule
\end{tabular}

\end{center}
\end{footnotesize}
\caption{A summary of the papers included in the study, with columns as follows. "Rank": the rank based on the published accuracy scores; "Key": reference to the paper; "$\overline{acc}$": the published accuracy score; "Num. image level": the number of image level figures shared by the authors; "Num. aggregated": the number of aggregated figures shared by the authors; "Annotator \#2 acc.": accuracy score of annotator \#2; "Region of eval.": the region of evaluation explicitly phrased by authors, 'F' for FoV, 'A' for all pixels, empty when it is not phrased explicitly; "Decimal places": the number of decimal places to which the figures are reported; "Citations": the number of citations at the time of writing this paper.}
\label{summary}
\end{table*}

A summary of the papers in Table \ref{summary} shows some interesting insights into the evaluation methodologies. 
Authors rarely phrase explicitly the region they use for evaluation, and when it is shared, a mixture of the two edge cases can be observed: the use of the FoV, and evaluation on all pixels. When the region of evaluation is not shared, one can still observe the two accuracy scores 0.9473 and 0.9637 for annotator \#2 (the scores we calculated in subsection \ref{ambig} under the FoV and using all pixels, repsectively). Finally, a slight variation can be observed in the scores being close to the one computed under the FoV (0.9473). This suggests that custom adjustments to the FoV may also have occured, but there is no textual evidence or indication of the nature of these adjustments.

\subsection{Overview of the numerical analysis}

The analysis is based on the interdependence of the scores $acc$, $sens$ and $spec$. Particularly, these scores cannot take arbitrary values: they are linear functions of the four unknowns $tp$, $tn$, $fp$ and $fn$, which need to sum up to the number of positive (vessel) $p = tp + fn$ and negative (non-vessel) $n = tn + fp$ pixels, while $p+n$ needs to match the size of the FoV mask or the size of the image. Consequently, the reported values of a triplet of $acc$, $sens$ and $spec$ together with the exact number of positive and negative pixels in the images must satisfy certain conditions. In the rest of the section, this idea is developed further to deal with the numerical uncertainties of rounding and averaging, and to allow inferences to be made about the evaluation methodologies.

\subsection{Image level analysis}

Some authors have reported the $acc$, $sens$ and $spec$ scores at the image level. In this subsection, we develop a consistency test to decide whether the image level scores were computed in the FoV region or using all pixels of the images.

\subsubsection{The image level consistency test}
\label{sec-img}

For each image, the true positives ($tp^*$) and true negatives ($tn^*$) calculated by the authors must satisfy the following inequalities by definition:
\begin{align}
sens - \epsilon &\leq \dfrac{tp^*}{p^*} \leq sens + \epsilon,\label{eq0}\\
spec - \epsilon &\leq \dfrac{tn^*}{n^*} \leq spec + \epsilon,\label{eq1}\\
acc - \epsilon &\leq \dfrac{tp^* + tn^*}{p^* + n^*} \leq acc + \epsilon,\label{eq2}\\
0 &\leq tp^* \leq p^*, \quad 0 \leq tn^* \leq n^*,\label{eq4}
\end{align}
where the total number of positive $p^*$ (vessel) and negative $n^*$ (non-vessel) pixels can be determined given the publicly available images, and $\epsilon$ denotes the numerical uncertainty of the reported figures. The setting $\epsilon = 10^{-k}/2$ is the maximum numerical uncertainty when the figures reported to $k$ decimal places are subject to rounding, and $\epsilon = 10^{-k}$ is the maximum uncertainty when truncation or ceiling to $k$ decimal places is assumed.

Eliminating $tp^*$ and $tn^*$ from the inequalities gives 6 conditions that must hold for a given triplet of $acc$, $sens$ and $spec$ scores, regardless of the actual values of $tp^*$ and $tn^*$:
\begin{align}
0 &\geq n^*(acc - spec) + p^*(acc - sens) - 2\epsilon(p^* + n^*), \nonumber\\
0 &\leq n^*(acc - spec) + p^*(acc - sens) + 2\epsilon(p^* + n^*), \nonumber\\
0 &\geq p^*(sens - \epsilon - 1), \quad 0 \leq p^*(sens + \epsilon),\nonumber\\
0 &\geq n^*(spec - \epsilon - 1), \quad 0 \leq n^*(spec + \epsilon). \label{il-cond}
\end{align}
The conditions (\ref{il-cond}) allow us to test the consistency of a certain triplet of $acc$, $sens$ and $spec$ reported for an image with $p^*$ positives and $n^*$ negatives, with $\epsilon$ numerical uncertainty. Since we are concerned about two cases (evaluation using the FoV mask or all pixels), one can extract the total number of positives and negatives with and without the FoV and test whether the conditions for either hypothesis are satisfied.
%Since we are concerned about two cases (using the FoV or using all pixels), for a particular image, one can extract the total number of positives and negatives with and without the FoV, and check if the conditions hold for any of the two hypotheses. 
For example, treating annotation \#1 as ground truth and annotation \#2 as a segmentation for the test image in Figure \ref{fig1}, the rounded performance scores with the FoV mask are $acc=0.9492$, $sens=0.7965$, $spec=0.9722$. In the FoV region, $p^*=29412$ and $n^*=194965$, and substituting these figures into the inequalities (\ref{il-cond}), all conditions are satisfied. On the other hand, including all pixels of the images, $p^*=29440$ and $n^*=300520$, and with this substitution the conditions fail.

The conditions (\ref{il-cond}) are sharp in the sense that if the assumptions on $p^*$ and $n^*$ are correct, then the conditions must hold; if the assumptions on $p^*$ and $n^*$ are sufficiently far the figures used to compute the scores $acc$, $sens$ and $spec$, then the test fails. However, due to the numerical uncertainty, scores computed in regions slightly different from the assumptions might also satisfy the conditions (\ref{il-cond}). This phenomenon is examined in the next subsection.

\begin{figure*}
\includegraphics[width=\textwidth]{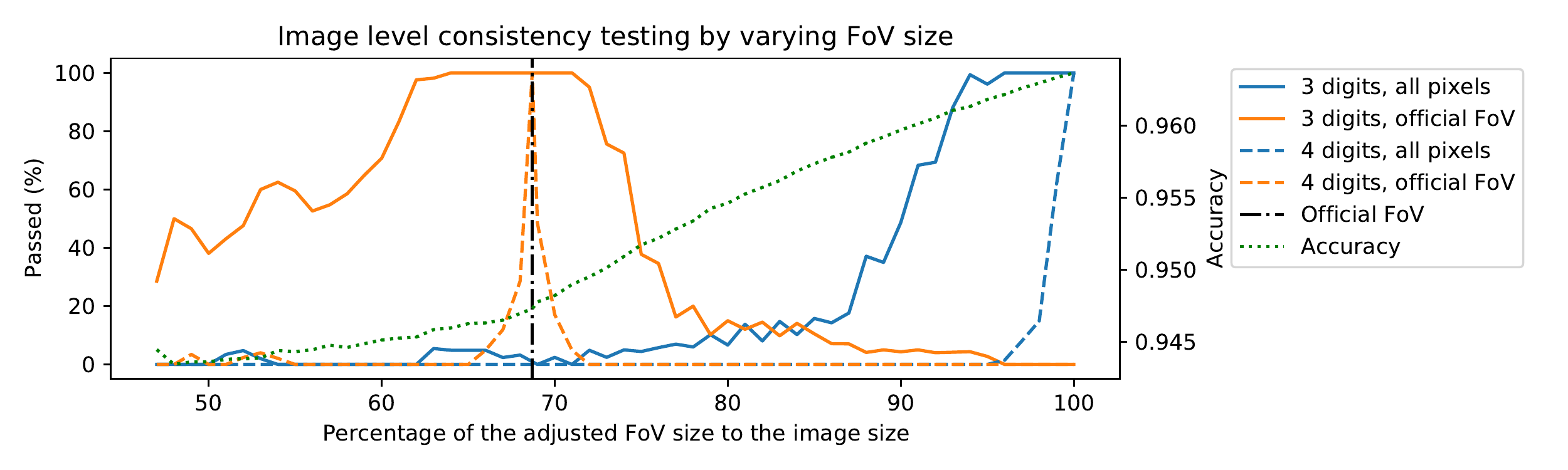}
\caption{The percentage of image level scores passing the consistency tests with the two assumptions when the region of evaluation and the rounding of performance scores is varied. %For a particular assumed evaluation methodology and rounding the curves represent the percentage of scores (test images) passing the tests. %When the scores are rounded to 4 digits, the window of acceptance decays rapidly as the size of the region deviates from the assumed one. The two corner cases are clearly distinguished by the test as only 0-5\% percents of the image level figures pass the test at the opposing hypotheses.
}
\label{sensitivity}
\end{figure*}

\subsubsection{Sensitivity of the image level consistency test}
 To quantitatively characterize the sensitivity of the test to changes in the region of evaluation, we carried out a simulation treating annotation \#1 as ground truth and annotation \#2 as segmentation for all 20 images in the DRIVE test set. The FoV masks were gradually dilated to cover the images; for each mask size, the performance scores were calculated and rounded to 3 and 4 digits; and the tests derived in subsection \ref{sec-img} were applied to the scores to check whether they were computed under the original FoV masks or using all pixels of the images. The percentages of image level scores passing the tests at certain mask sizes are plotted in Figure \ref{sensitivity} along with the average accuracy to illustrate the bias caused by the adjustment of evaluation region. The first thing to observe is that varying the size of the evaluation region has a remarkable effect on the accuracy scores. However, when the performance scores are rounded to 4 digits, a deviation as little as 1\% from the assumed region makes the test fail on more than 50\% of the 20 test images, as can be seen by the steep drop in the corresponding dashed lines around the size of the official FoVs (69\% of the image size) and at 100\% of the image size. The increased uncertainty of rounding to 3 digits widens the acceptance window (solid lines) and about 50\% of the scores pass despite a 15\% change in the size of the evaluation region. However, the passing rate drops to about 5\% at the opposite assumptions, indicating that the two edge cases can still be distinguished by the test. The analysis suggests that \emph{the consistency test is able to distinguish the cases where the official (or slightly differing) FoV masks were used from those cases when all (or almost all) pixels of the images were included in the evaluation}.

\begin{table}
\begin{center}
\begin{scriptsize}
\begin{tabular}{l@{\hspace{4pt}}l@{\hspace{4pt}}l@{\hspace{4pt}}l@{\hspace{4pt}}r@{\hspace{4pt}}r@{\hspace{4pt}}r@{\hspace{4pt}}r@{\hspace{4pt}}l@{\hspace{4pt}}}
\toprule
                            Key & $\overline{acc}$ & $\overline{sens}$ & $\overline{spec}$ &  \rotatebox{90}{Decimal places} &  \rotatebox{90}{Num. image level fig.} &  \rotatebox{90}{\% passed assum. FoV} &  \rotatebox{90}{\% passed assum. all pixels} &    Decision \\
\midrule
              \cite{hassan2018} &            .9793 &             .8981 &             .9883 &                               4 &                                     20 &                                                   0 &                                                   5 &     outlier \\
                \cite{wang2015} &            .9767 &             .8173 &             .9733 &                               4 &                                     20 &                                                   0 &                                                   0 &     outlier \\
          \cite{moghimirad2012} &            .9659 &             .7852 &             .9935 &                               4 &                                      4 &                                                  25 &                                                  25 &     outlier \\
  \cite{escorcia-gutierrez2020} &            .9640 &             .6170 &             .9980 &                               3 &                                     20 &                                                   0 &                                                   0 &     outlier \\
          \cite{narkthewan2019} &            .9617 &             .6392 &             .9920 &                               4 &                                     20 &                                                   0 &                                                   5 &     outlier \\
              \cite{waheed2015} &            .9616 &             .7937 &             .9779 &                               4 &                                     20 &                                                   0 &                                                  90 &  all pixels \\
                \cite{tang2017} &            .9611 &             .8174 &             .9747 &                               4 &                                     20 &                                                   0 &                                                  10 &     outlier \\
               \cite{tamim2020} &            .9607 &             .7542 &             .9843 &                               4 &                                     20 &                                                   0 &                                                   0 &     outlier \\
                 \cite{zhu2016} &            .9607 &             .7140 &             .9868 &                               4 &                                     20 &                                                   0 &                                                   0 &     outlier \\
           \cite{thangaraj2017} &            .9606 &             .8014 &             .9753 &                               4 &                                     20 &                                                   0 &                                                  10 &     outlier \\
             \cite{lupascu2016} &            .9606 &             .7006 &             .9857 &                               4 &                                     40 &                                                   0 &                                                  82 &  all pixels \\
             \cite{lupascu2010} &            .9597 &             .6728 &             .9874 &                               4 &                                      7 &                                                   0 &                                                  57 &  all pixels \\
               \cite{ricci2007} &            .9595 &             .7283 &             .9832 &                               4 &                                      2 &                                                   0 &                                                  50 &     outlier \\
               \cite{fathi2013} &            .9581 &             .7768 &             .9759 &                               4 &                                     20 &                                                   0 &                                                 100 &  all pixels \\
                \cite{dash2018} &            .9570 &             .7410 &             .9860 &                               3 &                                     20 &                                                  15 &                                                  10 &     outlier \\
              \cite{frucci2016} &            .9550 &             .6400 &             .9850 &                               3 &                                     20 &                                                   0 &                                                 100 &  all pixels \\
               \cite{saroj2020} &            .9544 &             .7307 &             .9761 &                               4 &                                     20 &                                                   0 &                                                 100 &  all pixels \\
        \cite{geetharamani2016} &            .9536 &             .7079 &             .9778 &                               4 &                                     20 &                                                   0 &                                                  85 &  all pixels \\
                \cite{meng2015} &            .9529 &             .7489 &             .9818 &                               4 &                                     20 &                                                   5 &                                                   0 &     outlier \\
                  \cite{li2016} &            .9527 &             .7569 &             .9816 &                               4 &                                      2 &                                                 100 &                                                   0 &         FoV \\
               \cite{imani2015} &            .9523 &             .7524 &             .9753 &                               4 &                                     20 &                                                   0 &                                                   0 &     outlier \\
               \cite{singh2016} &            .9522 &             .7594 &             .9708 &                               4 &                                     20 &                                                   0 &                                                 100 &  all pixels \\
                  \cite{mo2017} &            .9521 &             .7760 &             .9779 &                               4 &                                      2 &                                                 100 &                                                   0 &         FoV \\
                \cite{dash2020} &            .9520 &             .7560 &             .9810 &                               3 &                                     20 &                                                  25 &                                                  15 &     outlier \\
               \cite{singh2017} &            .9513 &             .7171 &             .9739 &                               4 &                                     20 &                                                   0 &                                                 100 &  all pixels \\
             \cite{bharkad2017} &            .9503 &             .7278 &             .9718 &                               4 &                                     20 &                                                   0 &                                                 100 &  all pixels \\
             \cite{barkana2017} &            .9502 &             .7224 &             .9840 &                               4 &                                      4 &                                                   0 &                                                   0 &     outlier \\
                \cite{khan2016} &            .9501 &             .7373 &             .9670 &                               4 &                                     20 &                                                   0 &                                                   5 &     outlier \\
               \cite{fraz2012b} &            .9480 &             .7406 &             .9807 &                               4 &                                      2 &                                                   0 &                                                   0 &     outlier \\
              \cite{rahebi2014} &            .9461 &             .7365 &             .9707 &                               4 &                                     20 &                                                   5 &                                                   0 &     outlier \\
               \cite{marin2011} &            .9452 &             .7067 &             .9801 &                               4 &                                     20 &                                                 100 &                                                   0 &         FoV \\
               \cite{adapa2020} &            .9450 &             .6994 &             .9811 &                               4 &                                     20 &                                                 100 &                                                   0 &         FoV \\
               \cite{kumar2020} &            .9432 &             .7503 &             .9717 &                               4 &                                     20 &                                                 100 &                                                   0 &         FoV \\
                \cite{fraz2012} &            .9430 &             .7152 &             .9769 &                               4 &                                      2 &                                                  50 &                                                   0 &     outlier \\
               \cite{emary2014} &            .9390 &             .7210 &             .9710 &                               3 &                                     20 &                                                 100 &                                                   0 &         FoV \\
          \cite{odstrcilik2013} &            .9340 &             .7060 &             .9693 &                               4 &                                     20 &                                                   0 &                                                   0 &     outlier \\
\bottomrule
\end{tabular}

\end{scriptsize}
\end{center}
\caption{Results of the image level consistency tests: the percentages indicate the fraction of image level scores passing the consistency test with a particular assumption on the region of evaluation.}
\label{imagelevelres}
\end{table}

\begin{figure}
     \begin{center}
     \includegraphics[width=0.5\textwidth]{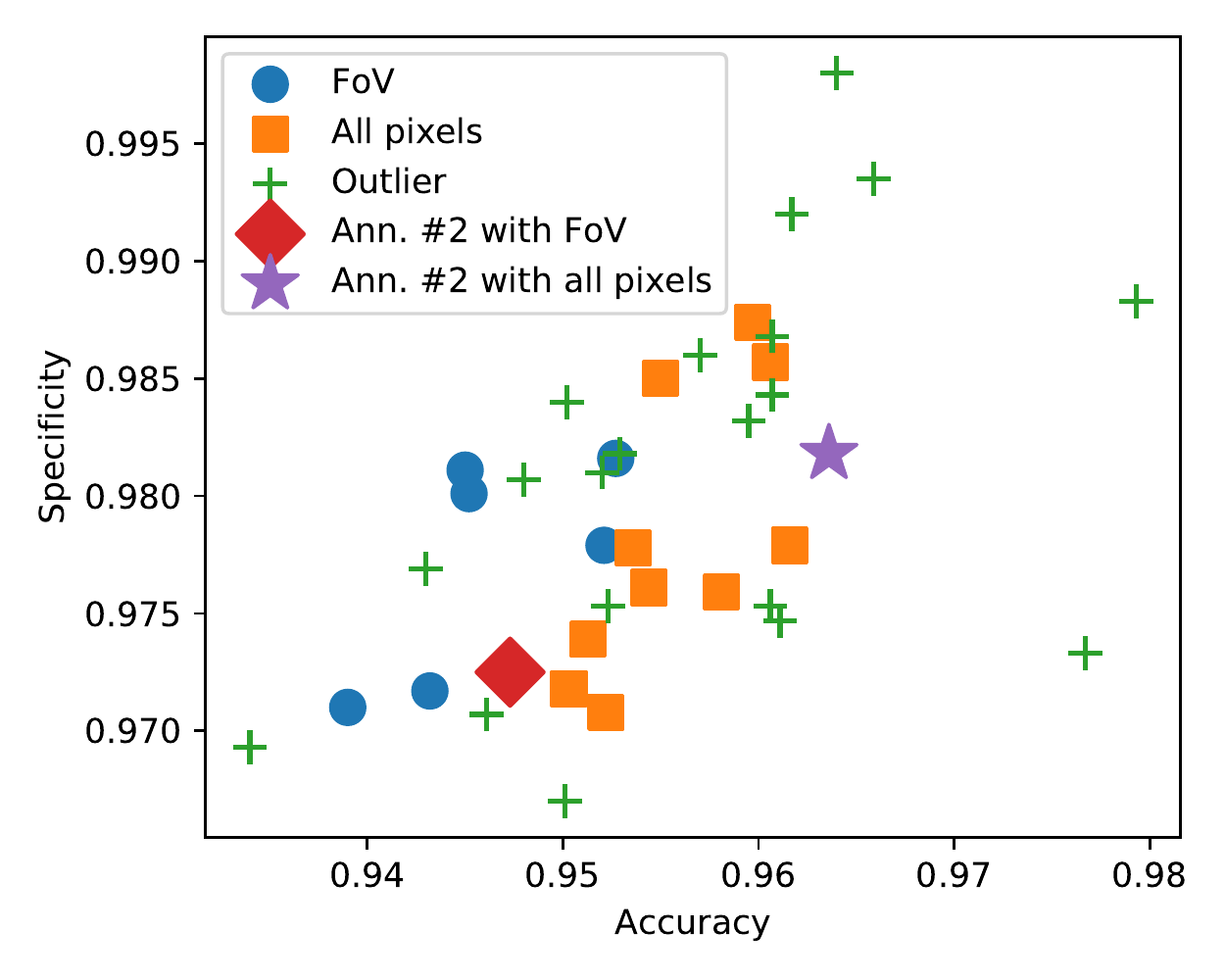}
     \end{center}
     \caption{The distribution of aggregated scores by the categories implied by image level figures.}
     \label{image-level-fig}
\end{figure}

\subsubsection{Image level results}

The image level consistency test was applied to all image level figures shared by 36 authors. 
To tolerate possible typos in the published figures and minor deviations from the exact assumptions, the image level results are aggregated according to the following permissive rule. If the use of the FoV (all pixels) is confirmed by the majority of the image level scores shared in a paper, the published scores are accepted to be calculated under the FoV (all pixels). If none of the assumptions are confirmed by the majority of the image level scores passing, all results of the paper are marked as \emph{outliers} in terms of evaluation: in these cases we can state with certainty that the authors evaluated in an unknown way. 

The results are summarized in Table \ref{imagelevelres} and plotted in Figure \ref{image-level-fig} in the accuracy-specificity plane. Surprisingly, 20 out of the 36 papers with image level scores turned out to be outliers. 
The ranking in Table \ref{imagelevelres} shows that almost all papers where the use of all pixels was accepted outperform those evaluating under the FoV mask, clearly showing the expected skew in the evaluations. A similar pattern can be seen in Figure \ref{image-level-fig}: the scores where the use of the FoV mask was accepted tend to have lower accuracy and specificity scores than those where the use of all pixels was confirmed. 

\subsection{The analysis of aggregated figures}

Most authors report only the averages of image level performance scores. In this subsection, we develop consistency tests to take into consideration the numerical uncertainty caused by the averaging.

\subsubsection{The consistency test for aggregated figures}
\label{sec-agg}

The concept of the consistency test for the average scores is similar to that of the image level test. Given the exact number of positive (vessel) and negative (non-vessel) pixels for each image in the DRIVE test set, one can formulate conditions that must be satisfied if the evaluation was performed according to the assumptions about the region of evaluation. Let $p_i^*, n_i^*, i=1, \dots, 20$ denote the number of positive and negative pixels in the manual annotation of the $i$th image under a particular assumption (the use of the FoV or all pixels). If the assumption is true, the following conditions must be satisfiable for some $tp_i$, $tn_i$, $i=1, \dots, 20$ unknown integers:

\begin{align}
\overline{sens} - \epsilon \leq \dfrac{1}{20}\sum\limits_{i=1}^{20}\dfrac{tp_i}{p_i^*} \leq \overline{sens} + \epsilon, \label{cond-sens}\\
\overline{spec} - \epsilon \leq \dfrac{1}{20}\sum\limits_{i=1}^{20}\dfrac{tn_i}{n_i^*} \leq \overline{spec} + \epsilon, \label{cond-spec}\\
\overline{acc} - \epsilon \leq \dfrac{1}{20}\sum\limits_{i=1}^{20}\dfrac{tp_i + tn_i}{p_i^* + n_i^*} \leq \overline{acc} + \epsilon, \label{cond-acc}\\
0 \leq tp_i \leq p_i^*, \quad 
0 \leq tn_i \leq n_i^*, \quad i=1, \dots, 20. \label{boundary-tn}
\end{align}
All these conditions are linear functions of the unknowns, thus, (\ref{cond-sens})-(\ref{boundary-tn}) can be interpreted as the condition set of a linear integer programming problem. 
In order to check if these conditions can be fulfilled, one can exploit any linear integer programming solver with a dummy objective function, and an empty feasibility set returned by the solver indicates that the conditions cannot be satisfied, thus, the aggregated performance scores are not consistent with the test images of DRIVE under the hypothesis (using the FoV or all pixels) which led to the $p_i^*$ and $n_i^*$ numbers.

\begin{figure*}
\includegraphics[width=\textwidth]{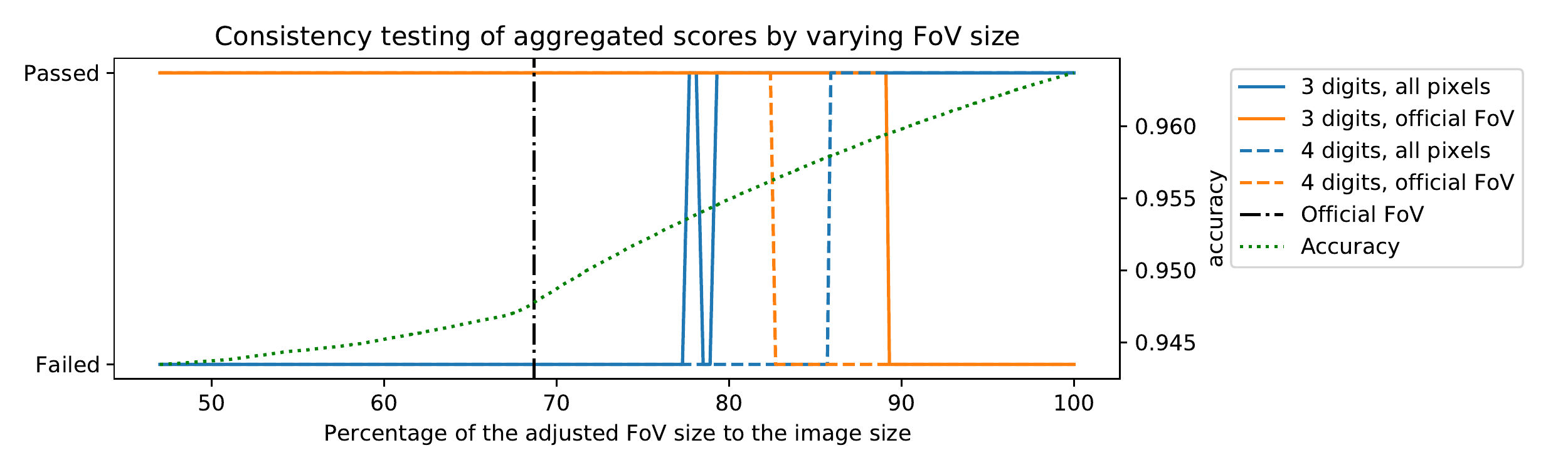}
\caption{Indicator curves for the aggregated scores showing whether the scores pass the consistency tests with the two assumptions when the region of evaluation and the rounding of performance scores are varied. 
}
\label{sensitivity-agg}
\end{figure*}

\begin{table}[htp]
\begin{scriptsize}
\begin{center}
\begin{tabular}{l@{\hspace{4pt}}l@{\hspace{4pt}}l@{\hspace{4pt}}l@{\hspace{4pt}}r@{\hspace{4pt}}r@{\hspace{4pt}}l@{\hspace{4pt}}l@{\hspace{4pt}}l@{\hspace{4pt}}}
\toprule
                             Key & $\overline{acc}$ & $\overline{sens}$ & $\overline{spec}$ &  \rotatebox{90}{Decimal places} &  \rotatebox{90}{Num. agg. figures} & \rotatebox{90}{Passed assum. FoV} & \rotatebox{90}{Passed assum. all pixels} &    Decision \\
\midrule
            \cite{jebaseeli2019} &            .9898 &             .8027 &             .9980 &                               4 &                                  1 &                                               &                                        &     outlier \\
  \cite{villalobos-castaldi2010} &            .9759 &             .9649 &             .9480 &                               4 &                                  1 &                                               &                                        &     outlier \\
               \cite{memari2017} &            .9722 &             .8726 &             .9884 &                               4 &                                  1 &                                               &                                        &     outlier \\
                 \cite{park2020} &            .9706 &             .8346 &             .9836 &                               4 &                                  1 &                                               &                                      + &  all pixels \\
                \cite{jiang2019} &            .9706 &             .8325 &             .9838 &                               4 &                                  1 &                                               &                                      + &  all pixels \\
                 \cite{atli2020} &            .9689 &             .7987 &             .9854 &                               4 &                                  1 &                                               &                                        &     outlier \\
                  \cite{pan2019} &            .9650 &             .8150 &             .9808 &                               4 &                                  1 &                                               &                                      + &  all pixels \\
             \cite{sreejini2015} &            .9633 &             .7132 &             .9866 &                               4 &                                  1 &                                               &                                      + &  all pixels \\
                \cite{saleh2011} &            .9630 &             .8423 &             .9658 &                               4 &                                  1 &                                               &                                        &     outlier \\
                \cite{kumar2016} &            .9626 &             .7006 &             .9871 &                               4 &                                  1 &                                               &                                      + &  all pixels \\
             \cite{wankhede2015} &            .9626 &             .7261 &             .9806 &                               4 &                                  1 &                                               &                                        &     outlier \\
               \cite{pandey2016} &            .9623 &             .8106 &             .9761 &                               4 &                                  1 &                                               &                                      + &  all pixels \\
                 \cite{alom2019} &            .9613 &             .7661 &             .9807 &                               4 &                                  1 &                                               &                                        &     outlier \\
                \cite{xiang2014} &            .9613 &             .7538 &             .9828 &                               4 &                                  1 &                                               &                                      + &  all pixels \\
               \cite{samuel2019} &            .9609 &             .8282 &             .9738 &                               4 &                                  1 &                                               &                                      + &  all pixels \\
                  \cite{fan2017} &            .9600 &             .7360 &             .9810 &                               3 &                                  1 &                                               &                                      + &  all pixels \\
                   \cite{wu2020} &            .9582 &             .7996 &             .9813 &                               4 &                                  1 &                                             + &                                        &         FoV \\
                \cite{budai2013} &            .9570 &             .6440 &             .9870 &                               3 &                                  1 &                                               &                                      + &  all pixels \\
                  \cite{noh2019} &            .9569 &             .8354 &             .9746 &                               4 &                                  1 &                                             + &                                        &         FoV \\
               \cite{soomro2019} &            .9560 &             .8700 &             .9850 &                               3 &                                  1 &                                               &                                        &     outlier \\
               \cite{frucci2017} &            .9560 &             .6600 &             .9850 &                               3 &                                  1 &                                               &                                      + &  all pixels \\
                  \cite{yan2018} &            .9542 &             .7653 &             .9818 &                               4 &                                  1 &                                             + &                                        &         FoV \\
                   \cite{na2018} &            .9540 &             .7680 &             .9700 &                               3 &                                  1 &                                               &                                      + &  all pixels \\
                 \cite{zhao2015} &            .9540 &             .7420 &             .9820 &                               3 &                                  1 &                                               &                                        &     outlier \\
                \cite{panda2016} &            .9539 &             .7328 &             .9752 &                               4 &                                  1 &                                               &                                      + &  all pixels \\
            \cite{liskowski2016} &            .9535 &             .7811 &             .9807 &                               4 &                                  1 &                                             + &                                        &         FoV \\
               \cite{soomro2018} &            .9534 &             .7592 &             .9763 &                               4 &                                  1 &                                               &                                        &     outlier \\
             \cite{dasgupta2017} &            .9533 &             .7691 &             .9801 &                               4 &                                  1 &                                             + &                                        &         FoV \\
                   \cite{hu2018} &            .9533 &             .7796 &             .9717 &                               4 &                                  1 &                                               &                                        &     outlier \\
                  \cite{ngo2017} &            .9533 &             .7464 &             .9836 &                               4 &                                  1 &                                             + &                                        &         FoV \\
                 \cite{yang2020} &            .9532 &             .7349 &             .9743 &                               4 &                                  1 &                                               &                                      + &  all pixels \\
                \cite{zhao2015b} &            .9530 &             .7440 &             .9780 &                               3 &                                  1 &                                               &                                        &     outlier \\
              \cite{rahmani2020} &            .9521 &             .7400 &             .9726 &                               4 &                                  1 &                                               &                                      + &  all pixels \\
            \cite{chalakkal2017} &            .9518 &             .7386 &             .9769 &                               4 &                                  1 &                                               &                                        &     outlier \\
               \cite{mapayi2015} &            .9511 &             .7313 &             .9724 &                               4 &                                  1 &                                               &                                      + &  all pixels \\
                \cite{zhang2018} &            .9504 &             .8723 &             .9618 &                               4 &                                  1 &                                             + &                                        &         FoV \\
                 \cite{song2017} &            .9499 &             .7501 &             .9795 &                               4 &                                  1 &                                             + &                                        &         FoV \\
         \cite{roychowdhury2015} &            .9494 &             .7395 &             .9782 &                               4 &                                  1 &                                             + &                                        &         FoV \\
               \cite{kovacs2016} &            .9494 &             .7450 &             .9793 &                               4 &                                  1 &                                             + &                                        &         FoV \\
             \cite{brancati2018} &            .9490 &             .7820 &             .9760 &                               3 &                                  1 &                                             + &                                        &         FoV \\
               \cite{nazari2013} &            .9481 &             .7112 &             .9716 &                               4 &                                  1 &                                               &                                      + &  all pixels \\
                 \cite{kaur2017} &            .9480 &             .8730 &             .9869 &                               4 &                                  1 &                                               &                                        &     outlier \\
            \cite{palanivel2020} &            .9480 &             .7375 &             .9788 &                               4 &                                  1 &                                             + &                                        &         FoV \\
                 \cite{shah2017} &            .9479 &             .7205 &             .9814 &                               4 &                                  1 &                                             + &                                        &         FoV \\
                \cite{zhang2016} &            .9476 &             .7743 &             .9725 &                               4 &                                  1 &                                             + &                                        &         FoV \\
               \cite{shukla2020} &            .9476 &             .7015 &             .9836 &                               4 &                                  1 &                                             + &                                        &         FoV \\
                \cite{cheng2014} &            .9474 &             .7252 &             .9798 &                               4 &                                  1 &                                             + &                                        &         FoV \\
                 \cite{zhou2017} &            .9469 &             .8078 &             .9674 &                               4 &                                  1 &                                             + &                                        &         FoV \\
            \cite{melinscak2015} &            .9466 &             .7276 &             .9785 &                               4 &                                  1 &                                             + &                                        &         FoV \\
               \cite{rezaee2017} &            .9463 &             .7189 &             .9793 &                               4 &                                  1 &                                             + &                                        &         FoV \\
             \cite{mendonca2006} &            .9463 &             .7315 &             .9781 &                               4 &                                  1 &                                             + &                                        &         FoV \\
                 \cite{miri2011} &            .9458 &             .7352 &             .9795 &                               4 &                                  1 &                                             + &                                        &         FoV \\
         \cite{strisciuglio2016} &            .9454 &             .7777 &             .9702 &                               4 &                                  1 &                                             + &                                        &         FoV \\
               \cite{javidi2017} &            .9450 &             .7201 &             .9702 &                               4 &                                  1 &                                               &                                      + &  all pixels \\
         \cite{strisciuglio2015} &            .9442 &             .7655 &             .9704 &                               4 &                                  1 &                                             + &                                        &         FoV \\
            \cite{azzopardi2014} &            .9442 &             .7655 &             .9704 &                               4 &                                  1 &                                             + &                                        &         FoV \\
                \cite{staal2004} &            .9441 &             .7750 &             .9725 &                               4 &                                  1 &                                             + &                                        &         FoV \\
                  \cite{you2011} &            .9434 &             .7410 &             .9751 &                               4 &                                  1 &                                             + &                                        &         FoV \\
               \cite{soomro2017} &            .9432 &             .7523 &             .9760 &                               4 &                                  1 &                                             + &                                        &         FoV \\
                  \cite{dai2015} &            .9418 &             .7359 &             .9720 &                               4 &                                  1 &                                             + &                                        &         FoV \\
             \cite{anzalone2008} &            .9418 &             .7286 &             .9810 &                               4 &                                  1 &                                               &                                        &     outlier \\
           \cite{dizdaroglu2012} &            .9412 &             .7181 &             .9743 &                               4 &                                  1 &                                               &                                        &     outlier \\
     \cite{salazar-gonzalez2014} &            .9412 &             .7512 &             .9684 &                               4 &                                  1 &                                             + &                                        &         FoV \\
                \cite{zhang2010} &            .9382 &             .7120 &             .9724 &                               4 &                                  1 &                                             + &                                        &         FoV \\
\bottomrule
\end{tabular}

\end{center}
\end{scriptsize}
\caption{The results of the consistency tests for aggregated figures, '+' denoting the cases passing the test with a particular assumption.}
\label{summary-agg}
\end{table}

\subsubsection{Sensitivity of the consistency test for aggregated figures}
Similarly to the image level analysis, the aggregated figures must pass the test if the assumptions are met. However, due to the increased numerical uncertainty, it is questionable whether the sensitivity of the test is high enough to distinguish the two corner cases we are concerned with. As before, a quantitative characterization of sensitivity can be obtained by varying the size of the evaluation region, but this time checking whether the aggregated figures pass the test with the $p_i^*$, $n_i^*$, $i=1,\dots,20$ numbers extracted according to the hypotheses. According to the results shown in Figure \ref{sensitivity-agg}, a 15\% deviation in size from the hypothesised region causes the test to fail when the scores are reported to 4 digits. With rounding to 3 digits, the two corner cases are still distinguished by the test. However, if the region of evaluation is between 77\%-87\% of the image size, the scores can pass the test with either assumption. 

\begin{figure}
     \begin{center}
     \includegraphics[width=0.5\textwidth]{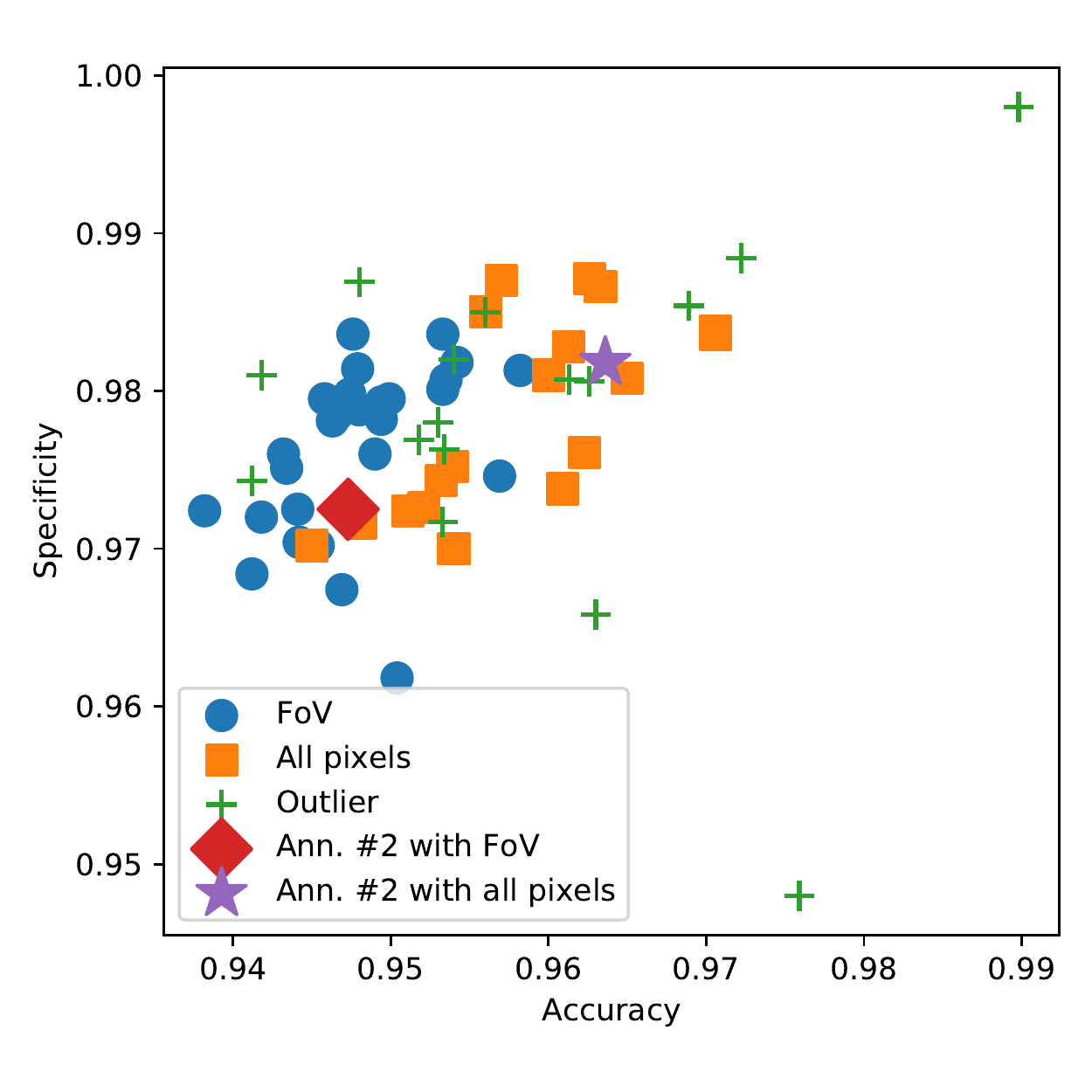}
     \end{center}
     \caption{The distribution of aggregated scores by the categories implied by the aggregated figures.}
     \label{aggregated-fig}
\end{figure}

\subsubsection{The results on aggregated figures}
In some papers (such as \citep{wu2020}), multiple triplets of aggregated scores are reported to illustrate the effects of the various steps of the proposed methods. Similarly to the image level analysis, the majority rule is applied to make clear decisions about the region of evaluation used by the authors. %Despite the sensitivity characteristics of the test allowing it, we did not find any case when both assumptions were accepted.
 The results are summarized in Table \ref{summary-agg} and plotted in Figure \ref{aggregated-fig}. Again, we assigned 3 labels to the papers, in 30 cases the use of the FoV was accepted, in 18 cases the use of all pixels was accepted, and 16 papers were classified as outliers (the scores do not pass the test with either of the hypothesised regions). As before, the ranking in Table \ref{summary-agg} and the illustration in Figure \ref{aggregated-fig} confirm the expected bias in the evaluations: the algorithms evaluated on all pixels tend to have higher accuracy and specificity scores, in general. 

To assess the sensitivity of the aggregated consistency test compared to the image level test, we compared the categorization based on image level and aggregated figures and found that the aggregated consistency test gave the same result in all cases where the use of the FoV mask or all pixels was confirmed at the image level. However, from the total of 20 outliers identified at the image level, in 15 cases the aggregated scores passed the consistency test with one of the assumptions. This suggests that there can be further outliers among the papers in which only aggregated scores are reported and the consistency test confirmed one of the hypotheses. 

\subsection{An improved ranking}
\label{sec-rank}

In this section, we attempt to derive a more realistic ranking of the algorithms by eliminating the bias caused by the differing regions of evaluation. The goal of the method we develop is to map the scores calculated on all pixels into the domain of evaluation with the FoV mask.

The idea is analogous to that of the consistency tests. Let the subscripts $x_F$ and $x_a$ refer to figures computed under the FoV mask and using all pixels, respectively, and let $x_d=x_a - x_F$ denote the difference. Given a particular triplet of image level scores $acc_a$, $sens_a$ and $spec_a$ reported by the authors with $\epsilon=10^{-k}$ numerical uncertainty, we would like to infer on $acc_F$, $sens_F$ and $spec_F$. Substituting the unknowns $tp_a^*= tp_F^* + tp_d^*$ and $tn_a^*=tn_F^* + tn_d^*$ into (\ref{eq0})-(\ref{eq4}), one gets the conditions
\begin{align}
sens_a - \epsilon &\leq \dfrac{tp_F^* + tp_d^*}{p_F^* + p_d^*} \leq sens_a + \epsilon \nonumber\\
spec_a - \epsilon &\leq \dfrac{tn_F^* + tn_d^*}{n_F^* + n_d^*} \leq spec_a + \epsilon,\nonumber\\
acc_a - \epsilon &\leq \dfrac{tp_F^* + tn_F^* + tp_d^* + tn_d^*}{p_F^* + n_F^* + p_d^* + n_d^*} \leq acc_a + \epsilon,\nonumber\\
tp_{F}^* \in \left[0, p_{F}^*\right], \quad
 tn_{F}^* &\in \left[0, n_{F}^*\right], \quad
 tp_{d}^* \in \left[0, p_{d}^*\right], \quad
 tn_{d}^* \in\left[0, n_{d}^*\right], \label{il-adjustment}
\end{align}
which must hold for some integer unknowns $tp_F^*, tn_F^*, tp_d^*$ and $tn_d^*$. The constants $p_F^*, p_d^*, n_F^*$ and $n_d^*$ are known from the publicly available test set. Focusing on the accuracy score $acc_{F}^*= \dfrac{tp_{F}^* + tn_{F}^*}{p_{F}^* + n_{F}^*}$ first, we are interested in a lower and upper bound on the values it can take under the conditions (\ref{il-adjustment}). Recognizing that the conditions (\ref{il-adjustment}) with the objective function $acc_F^*$ form a linear integer programming problem of the free integer variables $tp_F^*, tn_F^*, tp_d^*$ and $tn_d^*$, one can determine the minimum ($L$) and maximum ($U$) values the objective function can take by exploiting any linear programming solver. These extrema provide a sharp interval estimate for the accuracy score evaluated under the FoV: $acc_{F}^*\in[L(acc_{F}^*), U(acc_{F}^*)]$. The adjusted accuracy score is defined as the mid-point of the interval and the error of the adjustment is estimated as the half-width of the interval:
\begin{equation}
acc_F'= \dfrac{L(acc_{F}^*) + U(acc_{F}^*)}{2}, \quad \epsilon\left(acc_{F}'\right)= \dfrac{U(acc_{F}^*) - L(acc_{F}^*)}{2}.
\end{equation}
The adjusted image level $sens_F'$ and $spec_F'$ scores and corresponding error estimates are determined analogously. 

In practice, we found that the formulation (\ref{il-adjustment}) leads to wide intervals due to mathematical solutions that are not aligned with the problem. In the formulation (\ref{il-adjustment}), the number of true negatives outside the FoV ($tn_d$) can be zero, that is, the linear programming solver considers variable configurations corresponding to segmentations where all pixels outside the FoV are segmented as vessels, which is unlikely. %The interval estimates can be tightened
%which is unlikely. However, it is reasonable to assume that in some cases algorithms might overrun the boundary of the FoV and segment some negative background pixels as vessels. 
In order to tighten the intervals, the configuration space is reduced by making the natural assumption that no more negative pixels than 1\% of the vasculature are segmented as vessels outside the FoV: $tn_{d} \in\left[n_{d}^* - 0.01p_F^*, n_{d}^*\right]$. With this constraint, the worst-case error of the adjustment when applied to all image level figures reduces to $0.002$ with the average of 0.001, which we found sufficiently tight for further analysis. 
%We highlight that the improved constraint is favorable regarding the performance scores: the closer $tn_{d}$ is forced to $n_{d}$, the higher the adjusted scores are.

%\subsubsection{Adjustment of the aggregated figures}
%In many papers, only the averages of image level scores are shared, therefore . 
Similarly to the image level adjustment, for the averaged scores, one can substitute $tp_{a,i}^*= tp_{F,i}^* + tp_{d,i}^*$ and $tn_{a,i}^*= tn_{F,i}^* + tn_{d,i}^*$, $i=1,\dots,20$ into the conditions (\ref{cond-sens}) - (\ref{boundary-tn}) leading to a set of linear integer constraints. Then, minimizing and maximizing the objective function 
\begin{equation}
\overline{acc}_F^*= \sum\limits_{i=1}^{20}\dfrac{tp_{F,i}^* + tn_{F,i}^*}{p_{F,i}^* + n_{F,i}^*}
\end{equation}
in the free integer variables $tp_{F,i}^*, tn_{F,i}^*, tp_{d,i}^*, tn_{d,i}^*$, $i=1, \dots, 20$, one gets a sharp interval estimate for $\overline{acc}_F^*$. 
The adjusted aggregated score $\overline{acc_{F}}'$ is determined as the mid-point of the interval and the corresponding error term $\epsilon_{\overline{acc_{F}}}$ as the half-width of the interval. The adjusted $\overline{sens}_F^*$ and $\overline{spec}_F^*$ scores are determined analogously.

%The adjustment of aggregated scores is illustrated on the scores reported in a previous paper of ours \citep{kovacs2016}. 
%In order to illustrate the adjustment of the aggregated scores, we use those published in the previous papers of ours \citep{kovacs2016}, where we shared evaluation scores using the FoV and also using all pixels of the images. 
For illustration, the adjustment was applied to the scores calculated on all pixels and reported in our earlier paper \citep{kovacs2016}. The adjusted scores $\overline{acc}_{F}' = 0.9500$, $\overline{sens}_{F}' = 0.7447$, $\overline{spec}_{F}' = 0.9802$ are well aligned with the true scores $\overline{acc_{F}}=0.9494$, $\overline{sens_{F}}=0.7446$, $\overline{spec_F}=0.9793$ calculated in the FoV area, the error of the adjustment is not higher than $0.001$ for any of the figures.

The adjustment was applied to the reported scores of all methods we identified to be evaluated using all pixels. The worst case of the adjustment is $0.003$ with the average of $0.001$. As the errors are an order of magnitude smaller than the bias, we believe that the ranking based on the adjusted scores in Table \ref{ranking} is a more reliable baseline for the field than the original ranking in Table \ref{summary}. 

\begin{table}
\begin{scriptsize}
\begin{center}
\begin{tabular}{l@{\hspace{3pt}}l@{\hspace{3pt}}l@{\hspace{3pt}}l@{\hspace{3pt}}l@{\hspace{3pt}}l@{\hspace{3pt}}l@{\hspace{3pt}}l@{\hspace{3pt}}}
\toprule
                          Key & \rotatebox{90}{Rank} & \rotatebox{90}{Original rank} & \rotatebox{90}{Rank diff.} & \rotatebox{90}{Adjusted acc.} & \rotatebox{90}{Published acc.} &    Category &      Operation \\
\midrule
                \cite{wu2020} &                    1 &                            30 &                         29 &                         .9582 &                          .9582 &         FoV &  deep learning \\
              \cite{park2020} &                    2 &                             6 &                          4 &                         .9578 &                          .9706 &  all pixels &  deep learning \\
             \cite{jiang2019} &                    3 &                             7 &                          4 &                         .9578 &                          .9706 &  all pixels &  deep learning \\
               \cite{noh2019} &                    4 &                            34 &                         30 &                         .9569 &                          .9569 &         FoV &  deep learning \\
               \cite{yan2018} &                    5 &                            39 &                         34 &                         .9542 &                          .9542 &         FoV &  deep learning \\
         \cite{liskowski2016} &                    6 &                            44 &                         38 &                         .9535 &                          .9535 &         FoV &  deep learning \\
               \cite{ngo2017} &                    7 &                            47 &                         40 &                         .9533 &                          .9533 &         FoV &  deep learning \\
          \cite{dasgupta2017} &                    8 &                            46 &                         38 &                         .9533 &                          .9533 &         FoV &  deep learning \\
                \cite{li2016} &                    9 &                            52 &                         43 &                         .9527 &                          .9527 &         FoV &  deep learning \\
                \cite{mo2017} &                   10 &                            55 &                         45 &                         .9521 &                          .9521 &         FoV &  deep learning \\
               \cite{pan2019} &                   11 &                            10 &                         -1 &                         .9505 &                          .9650 &  all pixels &  deep learning \\
             \cite{zhang2018} &                   12 &                            61 &                         49 &                         .9504 &                          .9504 &         FoV &  deep learning \\
              \cite{song2017} &                   13 &                            65 &                         52 &                         .9499 &                          .9499 &         FoV &  deep learning \\
      \cite{roychowdhury2015} &                   14 &                            67 &                         53 &                         .9494 &                          .9494 &         FoV &      classical \\
            \cite{kovacs2016} &                   15 &                            66 &                         51 &                         .9494 &                          .9494 &         FoV &      classical \\
          \cite{brancati2018} &                   16 &                            68 &                         52 &                         .9490 &                          .9490 &         FoV &  deep learning \\
         \cite{palanivel2020} &                   17 &                            71 &                         54 &                         .9480 &                          .9480 &         FoV &     supervised \\
              \cite{shah2017} &                   18 &                            73 &                         55 &                         .9479 &                          .9479 &         FoV &     supervised \\
             \cite{zhang2016} &                   19 &                            74 &                         55 &                         .9476 &                          .9476 &         FoV &      classical \\
            \cite{shukla2020} &                   20 &                            75 &                         55 &                         .9476 &                          .9476 &         FoV &      classical \\
             \cite{cheng2014} &                   21 &                            76 &                         55 &                         .9474 &                          .9474 &         FoV &     supervised \\
          \cite{sreejini2015} &                   22 &                            12 &                        -10 &                         .9471 &                          .9633 &  all pixels &      classical \\
              \cite{zhou2017} &                   23 &                            77 &                         54 &                         .9469 &                          .9469 &         FoV &          other \\
         \cite{melinscak2015} &                   24 &                            78 &                         54 &                         .9466 &                          .9466 &         FoV &  deep learning \\
            \cite{rezaee2017} &                   25 &                            80 &                         55 &                         .9463 &                          .9463 &         FoV &      classical \\
          \cite{mendonca2006} &                   26 &                            79 &                         53 &                         .9463 &                          .9463 &         FoV &      classical \\
             \cite{kumar2016} &                   27 &                            14 &                        -13 &                         .9463 &                          .9626 &  all pixels &      classical \\
              \cite{miri2011} &                   28 &                            82 &                         54 &                         .9458 &                          .9458 &         FoV &      classical \\
            \cite{pandey2016} &                   29 &                            16 &                        -13 &                         .9458 &                          .9623 &  all pixels &      classical \\
      \cite{strisciuglio2016} &                   30 &                            83 &                         53 &                         .9454 &                          .9454 &         FoV &     supervised \\
             \cite{xiang2014} &                   31 &                            19 &                        -12 &                         .9452 &                          .9613 &  all pixels &      classical \\
             \cite{marin2011} &                   32 &                            84 &                         52 &                         .9452 &                          .9452 &         FoV &     supervised \\
             \cite{adapa2020} &                   33 &                            85 &                         52 &                         .9450 &                          .9450 &         FoV &     supervised \\
            \cite{waheed2015} &                   34 &                            18 &                        -16 &                         .9447 &                          .9616 &  all pixels &     supervised \\
         \cite{azzopardi2014} &                   35 &                            87 &                         52 &                         .9442 &                          .9442 &         FoV &      classical \\
      \cite{strisciuglio2015} &                   36 &                            88 &                         52 &                         .9442 &                          .9442 &         FoV &      classical \\
             \cite{staal2004} &                   37 &                            89 &                         52 &                         .9441 &                          .9441 &         FoV &     supervised \\
            \cite{samuel2019} &                   38 &                            22 &                        -16 &                         .9437 &                          .9609 &  all pixels &  deep learning \\
               \cite{you2011} &                   39 &                            90 &                         51 &                         .9434 &                          .9434 &         FoV &     supervised \\
           \cite{lupascu2016} &                   40 &                            26 &                        -14 &                         .9433 &                          .9606 &  all pixels &          other \\
            \cite{soomro2017} &                   41 &                            91 &                         50 &                         .9432 &                          .9432 &         FoV &      classical \\
             \cite{kumar2020} &                   42 &                            92 &                         50 &                         .9432 &                          .9432 &         FoV &      classical \\
               \cite{fan2017} &                   43 &                            27 &                        -16 &                         .9421 &                          .9600 &  all pixels &      classical \\
           \cite{lupascu2010} &                   44 &                            28 &                        -16 &                         .9421 &                          .9597 &  all pixels &     supervised \\
               \cite{dai2015} &                   45 &                            94 &                         49 &                         .9418 &                          .9418 &         FoV &      classical \\
  \cite{salazar-gonzalez2014} &                   46 &                            97 &                         51 &                         .9412 &                          .9412 &         FoV &          other \\
             \cite{fathi2013} &                   47 &                            31 &                        -16 &                         .9397 &                          .9581 &  all pixels &      classical \\
             \cite{emary2014} &                   48 &                            98 &                         50 &                         .9390 &                          .9390 &         FoV &      classical \\
             \cite{zhang2010} &                   49 &                            99 &                         50 &                         .9382 &                          .9382 &         FoV &      classical \\
             \cite{budai2013} &                   50 &                            33 &                        -17 &                         .9381 &                          .9570 &  all pixels &      classical \\
            \cite{frucci2017} &                   51 &                            36 &                        -15 &                         .9366 &                          .9560 &  all pixels &      classical \\
            \cite{frucci2016} &                   52 &                            37 &                        -15 &                         .9348 &                          .9550 &  all pixels &      classical \\
             \cite{saroj2020} &                   53 &                            38 &                        -15 &                         .9344 &                          .9544 &  all pixels &      classical \\
             \cite{panda2016} &                   54 &                            42 &                        -12 &                         .9336 &                          .9539 &  all pixels &      classical \\
      \cite{geetharamani2016} &                   55 &                            43 &                        -12 &                         .9330 &                          .9536 &  all pixels &     supervised \\
              \cite{yang2020} &                   56 &                            49 &                         -7 &                         .9325 &                          .9532 &  all pixels &      classical \\
                \cite{na2018} &                   57 &                            41 &                        -16 &                         .9325 &                          .9540 &  all pixels &      classical \\
             \cite{singh2016} &                   58 &                            54 &                         -4 &                         .9311 &                          .9522 &  all pixels &      classical \\
           \cite{rahmani2020} &                   59 &                            56 &                         -3 &                         .9309 &                          .9521 &  all pixels &      classical \\
             \cite{singh2017} &                   60 &                            59 &                         -1 &                         .9298 &                          .9513 &  all pixels &      classical \\
            \cite{mapayi2015} &                   61 &                            60 &                         -1 &                         .9295 &                          .9511 &  all pixels &      classical \\
           \cite{bharkad2017} &                   62 &                            62 &                          0 &                         .9284 &                          .9503 &  all pixels &      classical \\
            \cite{nazari2013} &                   63 &                            69 &                          6 &                         .9250 &                          .9481 &  all pixels &      classical \\
            \cite{javidi2017} &                   64 &                            86 &                         22 &                         .9231 &                          .9450 &  all pixels &     supervised \\
\bottomrule
\end{tabular}

\end{center}
\end{scriptsize}
\caption{The improved ranking based on adjusted performance scores, without outliers.}
\label{ranking}
\end{table}

\section{Insights}
\label{overall}

Based on the raw results and the adjusted scores, in this section, we provide some detailful insights into the field. 
%We highlight that all insights we phrase are based on a population of papers selected to represent influential papers in terms of citations. Other selections might lead to slightly different results.

\subsection{Insights from the consistency testing}
%\subsection{There is no standard evaluation methodology}
Overall, the use of the FoV mask was accepted in 36 cases, and the scores reported in 28 papers were consistent with the use of all pixels for evaluation. 36 papers were found to be outliers with undoubtful evidence that the evaluation was carried out in an unknown third way. Binomial hypothesis testing at the usual significance levels fails to reject the null hypothesis that the authors have no preference in the choice of the evaluation methodology (p-value: 0.19). Consequently, we can state that \emph{neither of the two evaluation methodologies we investigated can be treated as a standard in the field}.

%\subsection{Details of the evaluation methodology are rarely shared}
Explicit phrasing of the region of evaluation was found only in 29 papers. Out of them, only in 15 cases were the reported scores consistent with the claims of the authors. 
These inconsistencies indicate that \emph{the most reliable source for the region of evaluation is the triplet of performance scores the authors shared}.

%\subsection{Rankings are based on incomparable figures in 100+ papers}
The results of the consistency tests can be used to verify that the rankings in the papers are valid: if at least two algorithms evaluated in different regions are included in a ranking, it contains non-comparable figures.
%Whenever the scores reported in a paper do not pass a consistency test with a particular assumption on the region of evaluation, 
%Due to the nature of the consistency tests, papers to which different categories are assigned used different evaluation methodologies with certainty. With the categorization in hand, one can check if the rankings in the papers were valid: if at least two different categories are involved in a ranking, it contains non-comparable figures. 
This phenomenon was confirmed in 91 of the 100 papers. Together with the 10 reviews \citep{almotiri2018, moccia2018, mookiah2020, besenczi2016, khan2018, cetinkaya2020, lsrinidhi2017, singh2020, fraz2012-survey, fu2018} which only recite published figures, the statement in the title of this paper is confirmed: \emph{rankings are based on non-comparable figures in 100+ papers}. As recent papers are also affected, we can conclude that \emph{ranking algorithms based on non-comparable figures is an existing and ongoing problem in the field}.
%This phenomenon is confirmed in  \emph{the non-comparability of published figures is an existing and ongoing problem}.

\subsection{Insights from the new ranking}

The correlation coefficient between the original and the adjusted rankings is 0.18, indicating that \emph{the adjustment of scores leads to a substantial change in the recognition of efficient vessel segmentation techniques}.

The reported scores show that the accuracy of annotator \#2 was outperformed by 1\% in 2007, and at the time of writing, near-perfect techniques are available with accuracy of 0.98-0.99. In contrast, the adjusted scores show that the accuracy of annotator \#2 was only reached in 2014, and the highest accuracy to date is only 1\% above it (evaluated under the FoV mask). This phenomenon shows similarities with the field of electrohysterogram classification: the near perfect accuracy scores reported in the literature were caused by a methodological flaw in the evaluations \citep{vandewiele2020}.

\begin{figure*}
     \begin{center}
     \subfigure[]{\label{stackplot-0-a}\includegraphics[width=0.45\textwidth]{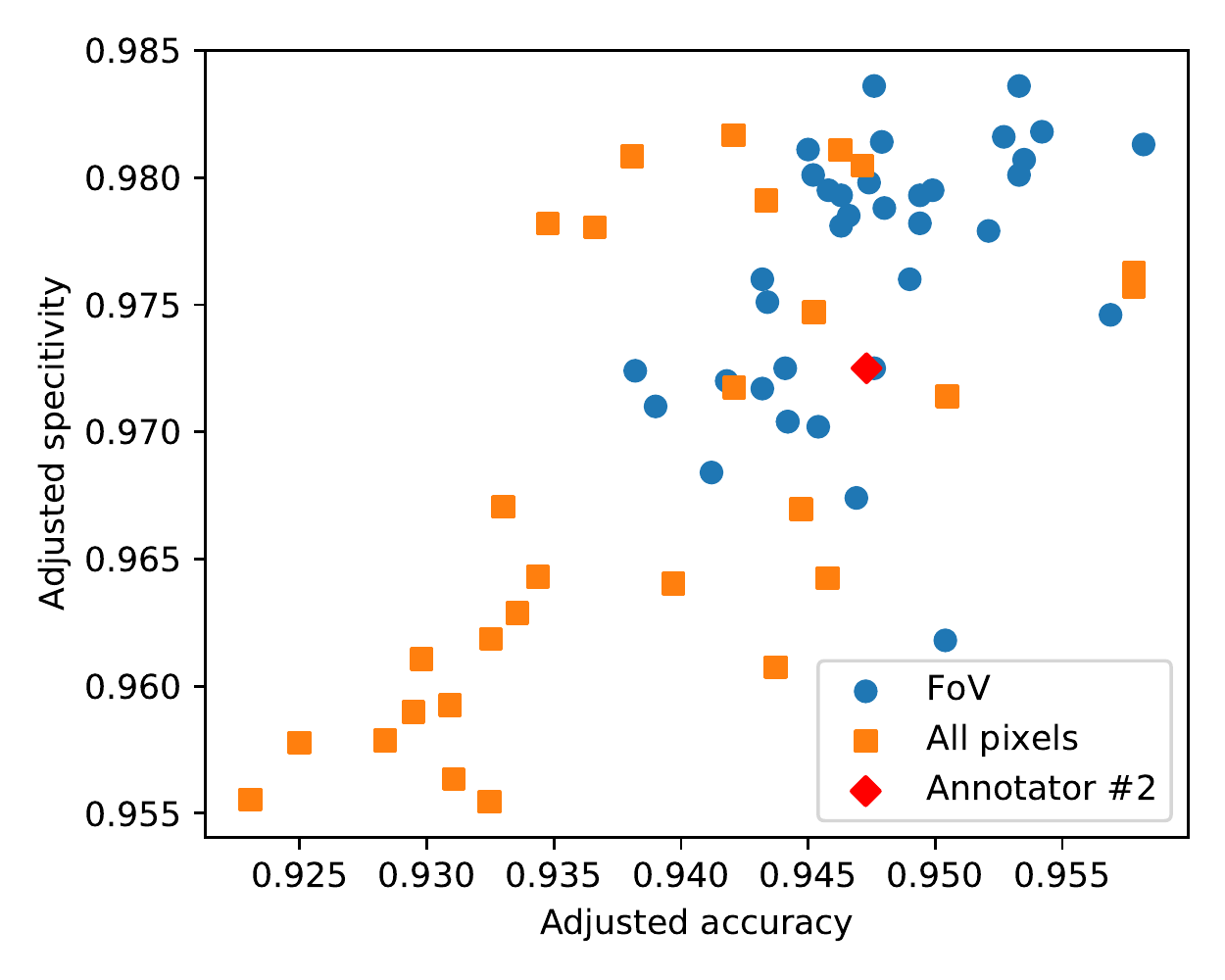}}
     \subfigure[]{\label{stackplot-0-b}\includegraphics[width=0.45\textwidth]{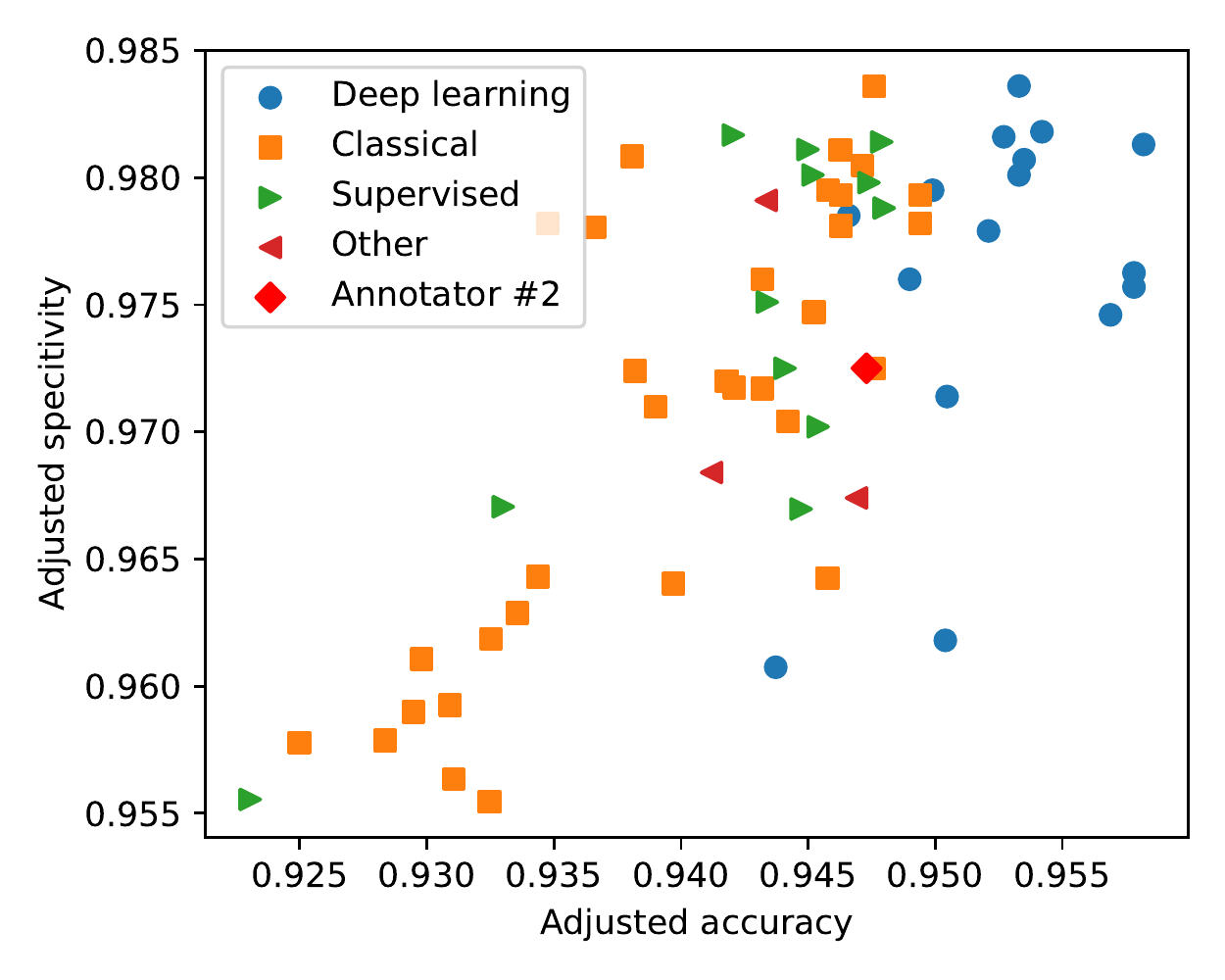}}
     \end{center}
     \caption{The distribution of the adjusted figures in the accuracy-specificity plane: by evaluation methodologies (a); and operating principles (b).}
     \label{stackplot-0}
\end{figure*}

In Figure \ref{stackplot-0-a} we have plotted the distribution of algorithms and evaluation methodologies in the adjusted accuracy-specificity plane. The plot shows that after eliminating the bias, \emph{the algorithms evaluated with all pixels have lower accuracy and specificity scores than those evaluated with the FoV mask.}
The independent two-sample t-test shows that the difference is statistically significant with a p-value of $10^{-9}$. The reason for this surprising phenomenon is that using all pixels to evaluate less effective techniques results in performance scores comparable to or even higher than those of more effective techniques evaluated under the FoV mask, and this brings publicity to less effective approaches. 

Finally, we compare the performances of the methods by their operating principles. We have introduced 4 categories: deep learning, classical (thresholding, filtering, morphology, region growing), supervised (feature extraction and supervised machine learning, but not deep learning), and others (dominantly graph-cut based segmentation techniques). The results presented in Figure \ref{stackplot-0-b} show that \emph{deep learning techniques increase the accuracy by almost 1\% compared to the best performing classical and supervised techniques}.

\subsection{A remark on the evaluation of deep learning techniques}

From a development perspective, there is an important difference between heuristic image processing pipelines and highly flexible deep learning solutions that calls into question the suitability of current evaluation methodologies in the field. Unlike the heuristic improvement of conventional image processing pipelines, the tuning of hyperparameters in deep neural networks is relatively straightforward (automated in many cases), which increases the risk of overfitting when the performance scores measured on the test set drive the tuning of the network. 

Although this could be prevented by using the training images of DRIVE for both training the weights in the network and tuning the network structure, we found evidence that some results are clearly overfitted to the test set: Both \citep{li2016} and \citep{liskowski2016} optimize the network size and hyperparameters on the test images. Despite the obvious overfitting, we did not discard these results for two reasons. First, all other papers using deep learning declare the network structures and the hyperparameters. None of the authors describe how the network was derived from the training data, which risks the overfitting of the network structure to the test images. On the other hand, there is no doubt that deep learning techniques achieve better results than classical approaches. Although some papers may have reported too optimistic scores due to overfitting to the test images, they clearly dominate the field. 

Consequently, the development and evaluation of deep learning solutions for retinal vessel segmentation should follow the standards of the deep learning literature by using 3 sets of images \citep{split}: one to train the weights in the network, another to tune the network structure and a final one to test the performance of the optimised network. However, since DRIVE specifies only 2 subsets and further splitting of the training set (20 images) would lead to a nearly insufficient number of images for deep learning, a larger set of images should become the standard in this field.

\subsection{Limitations}

The analysis we carried out has numerous limitations. Despite the systematic search for influential papers, the selected group may provide a biassed representation of the field and the results of the selection process may change over time as the number of citations changes.

The consistency tests for aggregated figures tolerate greater deviation from the hypotheses than do the image level tests. This inherent imbalance in the rigour of the tests implies that some papers that passed the consistency test with aggregated scores might have failed if the authors had reported image level scores.

The adjusted scores have some numerical uncertainty. Although the estimated errors are small, the improved ranking is also uncertain to this degree.

We have focused primarily on the comparability of the reported performance scores. However, higher performance is not the only goal in science. If some methods are easier to understand, maintain, adapt, or are more efficient in terms of runtime, they are still valuable contributions to science. Therefore, the improved ranking does not devalue the contributions of any paper, as all papers provide insights into the capabilities of the implemented approaches. 

\section{Summary and Conclusions}
\label{sec-discussion}

In this study, we investigated the coherence of reported performance scores in the segmentation of vessels in retinal images. We collected 100 influential papers reporting accuracy, sensitivity, and specificity for the publicly available data set DRIVE. Based on textual references to two corner cases of evaluation methodologies (evaluation under the FoV mask or using all pixels of the images), we developed numerical methods to test the consistency of the reported scores with the two assumed corner cases (subsections \ref{sec-img}, \ref{sec-agg}). Based on results of consistency tests, we attempted to remove bias and make the scores comparable to create a new baseline ranking (Section \ref{sec-rank}). Section \ref{overall} summarizes various findings and insights and also discusses the limitations of our analysis. The insights clearly show that there is no consensus in the field on the evaluation methodology. However, in most of the papers, including reviews of the field, the rankings and the objective evaluation of operating principles are biassed by the direct comparison of non-comparable figures. Based on the papers we selected and the adjustment we implemented, the current state of the art accuracy is 0.9582 (evaluated in the FoV region), which is about 1.1\% higher than the performance of annotator \#2. Despite the limitations of our method, we believe that the adjusted ranking in Table \ref{ranking} provides a more realistic view of the state of the art than any ranking based on the direct comparison of reported figures. 

The findings have numerous implications regarding the suitability and potential improvements of data sets, evaluation methodologies and reporting standards in the field:
\begin{itemize}[noitemsep, topsep=0pt]
    \item \emph{A novel data set that takes into account the changing requirements of the field is desirable} - potentially, this could emerge as a curated fusion of existing vessel segmentation data sets that provides predefined training, test and validation subsets for the reliable evaluation of deep learning techniques, and the sizes of the subsets should be based on well-established principles \citep{testsize}.
    \item The published average scores on DRIVE usually differ in the 3th-4th digits. Since these are the averages of only 20 image level scores, one can assume a relatively large variance, which makes the rankings ambigious. To investigate whether there were statistically significant improvements, \emph{the authors should provide image level figures} at least as supplementary material and explicitly report more details about the evaluation methodology (e.g., the region they use to calculate the scores). %A bigger and more diverse data set could also help to scatter the averaged scores and magnify the differences across algorithms.
    \item \emph{Authors should report the \emph{Dice score}} \citep{dice}, $dice = (2\cdot tp)/(2\cdot tp + fp + fn)$, which is a commonly accepted measure of segmentation performance. Since $dice$ is inherently free of the number of true negatives, the bias we study in this paper is naturally eliminated. Treating retinal vessel segmentation as a pixelwise binary classification problem, the AUC score (the standard performance measure for binary classification in machine learning) provides good insight into the trade-off between the sensitivity and specificity of segmentation algorithms - \emph{if applicable (the segmentation is based on pixelwise vesselness probabilities), authors should report AUC scores}.
    \item Maximizing some pixelwise performance measures of segmentation does not necessarily improve clinical applications. For example, an algorithm that provides an average accuracy of 0.95 with a small standard deviation may lead to a more reliable grading of disease severity than an algorithm with an average accuracy of 0.96 and a huge standard deviation, which means that it fails under certain conditions. Consequently, \emph{the evaluation of vessel segmentation methods should be more closely linked to clinical applications}.
\end{itemize} 

The numerical techniques we have developed can be easily adapted to other problems where similar methodological flaws may arise.
To allow full reproducibility and further analysis, all raw data, the implementation of the analysis and all results are shared in the GitHub repository \url{www.github.com/gykovacs/retinal_vessel_segmentation}.

\section*{Acknowledgments}
Gy\"orgy Kov\'acs would like to thank the support of the initiative 1LU-51P-4E.
On behalf of Attila Fazekas, this work was supported by the construction EFOP-3.6.3-VEKOP-16-2017-00002. The project was supported by the European Union, co-financed by the European Social Fund. 
%Acknowledgments should be inserted at the end of the paper, before the
%references, not as a footnote to the title. Use the unnumbered
%Acknowledgements Head style for the Acknowledgments heading.

%\section*{References}

%Please ensure that every reference cited in the text is also present in
%the reference list (and vice versa).

%\section*{\itshape Reference style}

%Text: All citations in the text should refer to:
%\begin{enumerate}
%\item Single author: the author's name (without initials, unless there
%is ambiguity) and the year of publication;
%\item Two authors: both authors' names and the year of publication;
%\item Three or more authors: first author's name followed by `et al.'
%and the year of publication.
%\end{enumerate}
%Citations may be made directly (or parenthetically). Groups of
%references should be listed first alphabetically, then chronologically.

%\nocite{*}

\bibliographystyle{model2-names}
\bibliography{refs}

%\section*{Supplementary Material}
%
%Supplementary material that may be helpful in the review process should
%be prepared and provided as a separate electronic file. That file can
%then be transformed into PDF format and submitted along with the
%manuscript and graphic files to the appropriate editorial office.

\end{document}